\documentclass[aps,prb,twocolumn,amsmath,amssymb,nofootinbib,superscriptaddress,floatfix,eqsecnum]{revtex4}

\usepackage[dvips]{color} 
\usepackage{graphicx}	
\usepackage{dcolumn} 	
\usepackage{bm}      	
\usepackage{longtable}
\usepackage{latexsym}
\usepackage{amsmath,amssymb}
\usepackage{color}
\DeclareMathOperator{\Tr}{Tr}
\DeclareMathOperator{\sgn}{sgn}
\newcommand{\dvex}[1]{d^{d}\bm{\mathrm{#1}}}
\newcommand{\vex}[1]{\bm{\mathrm{#1}}}
\newcommand{\ord}[1]{\bm{\mathit{O}}\left(#1\right)}
\newcommand{\sqord}[1]{\bm{\mathit{O}}\left[#1\right]}
\newcommand{\Lambdah}{\hat{\Lambda}}

\newcommand{\Qh}{\hat{Q}}
\newcommand{\Wh}{\hat{W}}
\newcommand{\NoOp}[1]{\bigl[\!\!\bigl[{#1}\bigr]\!\!\bigr]}

\newcommand{\tz}{\tilde{z}}
\newcommand{\tl}{\tilde{l}}

\usepackage{color}

\begin{document}

\title{
Termination of typical wavefunction multifractal spectra
at the Anderson metal-insulator transition: Field theory description
using the functional renormalization group
}

\author{Matthew S.\ Foster}
\email{psiborf@rci.rutgers.edu} 
\affiliation{Physics Department,
             Columbia University,
             New York,
	     NY 10027,
	     USA}
\affiliation{Department of Physics and Astronomy, 
	     Rutgers University, 
	     Piscataway, 
	     NJ 08854, 
	     USA}
\author{Shinsei Ryu}
\affiliation{Department of Physics,
	     University of California,
	     Berkeley,
	     CA 94720,
	     USA}
\author{Andreas W.\ W.\ Ludwig}
\affiliation{Department of Physics,
             University of California, 
             Santa Barbara, 
             CA 93106,
	     USA}
\date{\today}

\begin{abstract}
We revisit the problem of wavefunction statistics at the Anderson
metal-insulator transition (MIT) of non-interacting electrons
in $d > 2$ spatial dimensions.
At the transition, the
complex
spatial structure of the critical wavefunctions
is reflected in the non-linear behavior of the multifractal spectrum of
generalized inverse participation ratios (IPRs).
Beyond the crossover from narrow to broad IPR statistics, 
which always occurs for sufficiently large moments of the wavefunction 
amplitude,
the spectrum obtained from a \textit{typical} wavefunction 
associated with a particular disorder realization 
differs markedly from that 
obtained from the \textit{disorder-averaged} IPRs.
This phenomenon is known as
the termination of the multifractal spectrum.
We provide a 
field theoretical
derivation for the termination of the typical multifractal spectrum,
by 
combining
the non-linear sigma model framework, conventionally used to
access the MIT in $d = 2 + \epsilon$ dimensions, with a
functional renormalization group (FRG) technique.
The FRG method deployed here was originally pioneered to study 
the properties of
the two-dimensional (2D) random phase XY model
[D.\ Carpentier and P.\ Le Doussal, Nucl.\ Phys.\ B \textbf{588}, 565 (2000)]. 
The same method
was
used to demonstrate the termination of the multifractal 
spectrum in 
the very special problem of 2D Dirac fermions subject to a random Abelian
vector potential. 
Our result shows that the {\it typical} multifractal wavefunction
spectrum and its termination can be obtained
at a generic Anderson localization transition in $d>2$,  within the 
standard field theoretical framework of the non-linear sigma model, when
combined with the FRG.
\end{abstract}

\maketitle


\section{Introduction \label{Intro}}

Quantum interference induced by multiple elastic impurity 
scattering can produce very complex spatial fluctuations 
in electronic wavefunctions. The statistics of these fluctuations
may be used to distinguish different regimes of qualitative 
wavefunction behavior, e.g.\ localized versus extended. Of 
particular interest are the wavefunction statistics at a 
delocalization transition, such as the Anderson metal-insulator 
transition (MIT)\cite{LeeRamakrishnan} at the mobility edge in three spatial 
dimensions,\cite{MirlinReview1and2,MirlinReview3} 
or the integer quantum Hall plateau (IQHP) transition in two dimensions.\cite{Janssen94,Huckestein95,MirlinReview3} 
Here, the spatial structure of the critical wavefunctions is known not to be 
characterized 
by just  a single (or a few) independent exponent(s), 
but by an infinite set
thereof (`multifractality'). More precisely, wavefunction statistics 
are encoded through the $\tau(q)$ spectrum, or its Legendre transform, 
the singularity spectrum 
$f(\alpha)$.\cite{Janssen94,Huckestein95,Efetov97,MirlinReview1and2,MirlinReview3,Surfacemultifractality}

The $\tau(q)$ spectrum is defined via the 
(generalized) inverse participation ratio (IPR),\cite{Wegner80} given by
\begin{eqnarray}\label{IPRDef}
	P_{q}(\varepsilon_{i}) \equiv
	\int_{L^{d}} \dvex{r}\,
	|\psi_{i}(\vex{r})|^{2q},
	\quad
	P_{q} \sim L^{-\tau(q)},
\end{eqnarray}
where $d$ is the spatial dimensionality of the system, $L^{d}$ denotes the system 
volume, and $|\psi_{i}(\vex{r})|^2$ is the probability density of a normalized eigenstate 
wavefunction $\psi_{i}(\vex{r})$ with energy $\varepsilon_{i}$, evaluated at the point $\vex{r}$.
For eigenenergies $\varepsilon$ lying within a band of extended plane wave states, 
$\tau(q)=d(q-1)$, 
while exponentially localized states  yield
$\tau(q) \sim 0$ for $L \gg \xi$, with $\xi$ the localization length.
Multifractal behavior refers to non-linear $q$-dependence of the $\tau(q)$ spectrum,
and occurs, e.g., at the mobility edge $\varepsilon = \varepsilon_{c}$ in a disordered
three-dimensional (3D) system of non-interacting electrons.\cite{CastellaniPeliti86}
The singularity spectrum $f(\alpha)$ is related to the $\tau(q)$ spectrum 
through the Legendre transformation,
\begin{eqnarray}\label{fDef}
	f(\alpha) = q \alpha - \tau(q),
	\quad
	\frac{d \tau(q)}{dq}=\alpha.
\end{eqnarray}
The set of points at which an eigenfunction takes the value
$|\psi(\vex{r})|^2 \sim L^{-\alpha}$
is distributed according to the weight $L^{f(\alpha)}$;\cite{Halsey86,Janssen94}
in this sense, the singularity spectrum characterizes the interwoven
fractal measures of the sample associated with differently-scaling 
components of wavefunction intensity.
The wavefunction statistics have been studied experimentally using 
thin microwave 
cavities;\cite{Kudrolli95}
a very broad distribution of the wavefunction intensity, indicative 
of multifractal behavior, was indeed observed.\cite{Fyodorov95,MirlinReview1and2,Efetov97}

The multifractal spectrum [$\tau(q)$ or $f(\alpha)$]
at a delocalization critical point
is 
universal,
and thus serves 
as 
a ``fingerprint'' 
of the spatial structure of wavefunctions.
Spectra have been computed numerically at myriad 
delocalization transitions occuring in various spatial dimensions; see e.g.\ 
Refs.~\onlinecite{SchreiberGrussbach91,PookJanssen91,Janssen94,Huckestein95,EversMirlin00,Evers01,Mildenberger01,Evers2008,Obuse2008,VasquezI2008,VasquezII2008}. 
In particular, extensive numerical studies of the IQHP transition\cite{PookJanssen91,Huckestein95,Evers01} 
employing different microscopic models have convincingly established the universality of the entire
$f(\alpha)$ spectrum. Recent work includes that of Refs.\ 
\onlinecite{Obuse2008} and \onlinecite{Evers2008},
which aim in part at decrypting the critical (conformal field) theory describing
the plateau transition.

To compute the
entire
multifractal spectrum analytically is, however,
a very  difficult task in generic systems. This is even more so
because
it is a non-analytic function of $q$ or $\alpha$.
As emphasized in 
Refs.~\onlinecite{Chamon96}, \onlinecite{MirlinEvers00}, and \onlinecite{MirlinReview3},
this non-analyticity is related to the fact that
the $\tau(q)$ and $f(\alpha)$ spectra are 
defined for a \textit{typical} 
representative wavefunction,
drawn in principle from a system 
in a single, fixed realization 
of the static disorder.
On the contrary,
analytical methods (i.e., those based upon field theories)
are best suited for calculating quenched \textit{averaged} quantities.
To be precise, we 
define,
following Ref.~\onlinecite{MirlinEvers00},
\emph{two} sets of multifractal 
statistics in terms of the IPR defined in Eq.~(\ref{IPRDef}):
\begin{subequations}\label{typvsavg}
\begin{align} 
	\tau(q) &
	\equiv 
	-\frac{d \, \overline{\ln P_{q}}}{d \ln L},
	\label{tautyp}
	\\
	\tilde{\tau}(q) &
	\equiv 
	-\frac{d \, \ln \overline{P_{q}}}{d \ln L}. 
	\label{tauavg}
\end{align}
\end{subequations}
In this equation, the overbar $\overline{\cdots}$
represents an average over 
realizations of the quenched disorder. 
The typical $\tau(q)$ spectrum in Eq.~(\ref{IPRDef}) obtains from the log of the IPR
for a representative wavefunction; since the latter quantity 
is 
expected
to be self-averaging at the 
delocalization transition,\cite{EversMirlin00,MirlinEvers00} we may introduce
an additional, though redundant ensemble average over disorder realizations, as in 
Eq.~(\ref{tautyp}).  
We have also defined $\tilde{\tau}(q)$ in Eq.~(\ref{tauavg}), which
obtains from the average of the IPR itself.
The averaged  IPR can be encoded through the moments
of the local density of states (LDOS) operator in
an effective low-energy field theory (see Sec.~\ref{sec: definitions and model},
below); then, the scaling dimensions of the LDOS moment operators directly determine
$\tilde{\tau}(q)$. 
No such effectively local construction exists for the 
typical spectrum $\tau(q)$, 
and in fact
``non-local'' (or more precisely,
``multilocal'') 
correlations play an essential role\cite{Kogan96,Carpentier00,Carpentier01,Mudry03}
in the ``termination'' (defined below) of the typical $\tau(q)$, as we show in this paper.

For not too large $|q|$, one expects that
\begin{equation}\label{TypISAvg}
	\tau(q) = \tilde{\tau}(q),
\end{equation}
which is the case when the IPR $P_{q}$ represents a
self-averaging quantity 
[see Subsection (\ref{ComparisonOtheMethods}) for a review].
At sufficiently large $|q|$, however,
$P_{q}$ becomes broadly 
distributed,\cite{Fyodorov95,Chamon96,PrigodinAltshuler98,MirlinReview1and2,EversMirlin00,MirlinEvers00} 
and the corresponding $\tilde{\tau}(q)$ spectrum, dominated now by 
``rare events''
induced by the disorder averaging procedure, 
deviates from $\tau(q)$.\cite{footnote-REM}
While $\tilde{\tau}(q)$ is always easier to evaluate
analytically, it is $\tau(q)$ that is 
most easily obtained from 
a representative wavefunction in numerics.\cite{Janssen94,Huckestein95}
By comparison, the \emph{average} $\tilde{\tau}(q)$ and $\tilde{f}(\alpha)$ spectra 
were computed only recently via numerics at the IQHP\cite{Evers01} and 
Anderson\cite{Mildenberger01,VasquezII2008} transitions.
\cite{AdditionalTermination}

In this paper we calculate the typical multifractal spectrum
at the Anderson MIT in the unitary\cite{LeeRamakrishnan} [broken time-reversal] 
symmetry
class of disordered, normal metals, in $d > 2$.
The spectrum $\tilde{\tau}(q)$ associated to the averaged IPR,
evaluated at the metal-insulator transition in $d = 2+\epsilon$,
was obtained long ago\cite{Wegner80,Pruisken85,HofWegner86,Wegner87} via standard perturbative renormalization 
group (RG). The form of the \emph{typical} $\tau(q)$ has been argued before only
on heuristic grounds.\cite{MirlinEvers00,MirlinReview3} 
We compute here for the first time the typical spectrum directly, 
using an (analytical)  functional renormalization group (FRG)
scheme\cite{Carpentier00,Carpentier01,Mudry03} previously employed in the study
of wavefunctions statistics in a 
special class of disordered Dirac fermion models in 
2D.\cite{Ludwig94,Mudry96,Kogan96,Chamon96,Castillo97,Carpentier00,Carpentier01,Motrunich02,Mudry03,
Fukui02,Yamada03,DellAnna2006}

\subsection{Average vs.\ typical spectra and termination}\label{AvgTypTerm}

In the 
field theory description
of Anderson localization
[especially the non-linear sigma model (NL$\sigma$M) 
formulation,\cite{Wegner79,LeeRamakrishnan} reviewed 
in Sec.~\ref{sec: definitions and model}]
the exponent $\tilde{\tau}(q)$, $q \in \mathbb{N}$ of
the averaged IPR
can be read off from the 
scaling dimensions $x^*_q$ and $x^*_{1}$ of local composite operators 
$\mathcal{O}_{q}(\vex{r})$ and $\mathcal{O}_{1}(\vex{r})$, which
represent the $q^{\mathrm{th}}$ and $1^{\mathrm{st}}$ moments of 
the local density of states (LDOS), 
respectively:\cite{Wegner80,AltshulerKratsovLerner86-89,AltshulerKratsovLerner91} 
\begin{equation}\label{AvgMFCSpecI}
	\tilde{\tau}(q)
	=
	d(q-1) + x^*_q - q \, x^*_{1}.
\end{equation}
(See Sec.~\ref{sec: definitions and model} for details.) 
For example, at the 
Anderson metal-insulator transition in $d = 2+\epsilon$ 
dimensions in the unitary symmetry class,
one obtains\cite{Pruisken85,HofWegner86,Wegner87}
\begin{subequations}\label{AvgMFCSpecII}
\begin{align}
	x^*_q 
	&=
	- \Xi \,
	q(q-1)
	+\sqord{\epsilon^2 q^2 (q-1)^2},
	\label{AvgMFCSpecxq}\\
	x^*_1 
	&=
	0,
	\label{AvgMFCSpecx1}\\
	\Xi 
	&= 
	\sqrt{\epsilon/2} 
	+\ord{\epsilon^{5/2}}
	\label{XiDef}.
\end{align}
\end{subequations}
We can define a corresponding \emph{average} singularity spectrum via
\begin{align}\label{AvgfSpec}
	\tilde{f}(\alpha)
	&\equiv
	q \alpha - \tilde{\tau}(q),
	\quad
	\frac{d \tilde{\tau}(q)}{d q}=\alpha,
	\nonumber\\
	&=
	d 
	-
	\tilde{f}_{2}
	\left(\alpha-\alpha_{0}\right)^2
	+
	\sqord{\sqrt{\epsilon}\left(\alpha-\alpha_{0}\right)^3},
\end{align}
where
\begin{subequations}
\begin{align}
	\tilde{f}_{2} 
	&= \frac{1}{4 \Xi} + \ord{\epsilon},
	\label{AvgfSpecCoeff}\\
	\alpha_{0} 
	&= d + \Xi + \ord{\epsilon^{5/2}}.
	\label{AvgfSpecMax}
\end{align}
\end{subequations}
The corrections to [$\ord{\cdots}$ terms in] Eqs.~(\ref{AvgMFCSpecxq}), (\ref{XiDef}), and 
(\ref{AvgfSpec})--(\ref{AvgfSpecMax})  
obtain at the fourth loop order\cite{Wegner87} (or beyond)
in the epsilon 
expansion.
By contrast, Eq.~(\ref{AvgMFCSpecx1}) is exact, and is equivalent to the 
statement that the average (global) density of states
is non-critical at the MIT in the unitary symmetry class.\cite{Wegner81,McKaneStone81,x1For2DchiralDirac}
In the present paper, we work only to the lowest non-trivial 
order in  the expansion parameter $\sqrt{\epsilon}$.
The consistency of the $\epsilon$-expansion
in dealing with high moments of the LDOS operator
is demonstrated in Sec.~\ref{sec: conclusion}.
Results similar to 
Eqs.~(\ref{AvgMFCSpecxq})--(\ref{XiDef})
were first computed for the 
time-reversal invariant orthogonal\cite{Wegner80,symplecticClass} symmetry class.
The so-obtained $\tilde{f}(\alpha)$ spectrum is consistent with large-scale 
numerics.\cite{Mildenberger01}

If one were to reconstruct 
the
probability
distribution of the wavefunction 
amplitudes
from the
average spectra [Eqs.~(\ref{AvgMFCSpecI}) and (\ref{AvgfSpec})],
a quadratic $\alpha$-dependence of $\tilde{f}(\alpha)$ 
implies log-normal asymptotics of the distribution function.
\cite{AltshulerKratsovLerner91,AltshulerKratsovLerner86-89,Muzykantskii94,Mirlin95,Falko95}
The precursor of this broad distribution is already visible
at the crossover
from the ballistic to diffusive regime,
where wavefunctions start to show (weak) Anderson localization.
\cite{LeeRamakrishnan}
In this 
``pre-localized'' regime,
renormalization group studies of an (extended) NL$\sigma$M,\cite{AltshulerKratsovLerner91,AltshulerKratsovLerner86-89}
as well as semi-classical analyses of the supersymmetric (SUSY) NL$\sigma$M\cite{Muzykantskii94,Falko95,Mirlin95}
predict that the distribution of the wavefunction amplitudes
starts to deviate from the Gaussian, developing a 
log-normal tail.\cite{Efetov97, AltshulerKratsovLerner91}
As it obtains from the $\tilde{\tau}(q)$ spectrum associated with the 
average of the IPR,\cite{AltshulerKratsovLerner91,AltshulerKratsovLerner86-89} 
this tail reflects the influence of rare realizations of the disorder
and so-called ``anomalously localized states.''\cite{Muzykantskii94,Falko95,Mirlin95}
Even though the tail of the distribution is still small, describing
rare events in the mesoscopic regime, it is responsible for 
anomalous current relaxation, which is slower than expected from the Drude formula.\cite{AltshulerKratsovLerner91,Muzykantskii94}

For small $\sqrt{\epsilon}$ (i.e., weak disorder) and $q$ not too large,
one might be inclined to expect that the results for $\tilde{\tau}(q)$ and 
$\tilde{f}(\alpha)$ in Eqs.~(\ref{AvgMFCSpecI})--(\ref{AvgfSpecMax}) should
not differ substantially from $\tau(q)$ and $f(\alpha)$, respectively.
However, the range of applicability of Eq.~(\ref{AvgfSpec}) to
the typical $f(\alpha)$ is limited to $\alpha_{-} \leq \alpha \leq \alpha_{+}$,
where, to lowest order
\begin{equation}\label{alphapmDef}
	\alpha_{\pm}
	\equiv
	\left(\sqrt{d} \pm \sqrt{\Xi}\right)^2 
	+ \ldots,
\end{equation}
so that $f(\alpha_{\pm}) = 0$.
For $\alpha > \alpha_{+}$ and $\alpha < \alpha_{-}$,
the average singularity spectrum $\tilde{f}(\alpha)$ becomes negative,
which does not make sense 
if it is interpreted for a 
typical wavefunction [see the discussion following 
Eq.~(\ref{fDef}), above].
These thresholds define 
the critical values $q_c^{\pm}$ of $q$ 
for the $\tau(q)$ spectrum through
\begin{align}\label{pmqcDef}
        q_{c}^{\pm} 
	&\equiv 
	\frac{d\tilde{f}(\alpha_{\mp})}{d\alpha} = \pm q_c + 
	\ldots,
	\nonumber\\
	q_c&
	=
        \sqrt{
        \frac{d}{\Xi}}.
\end{align}

\begin{figure}
\includegraphics[width=0.4\textwidth]{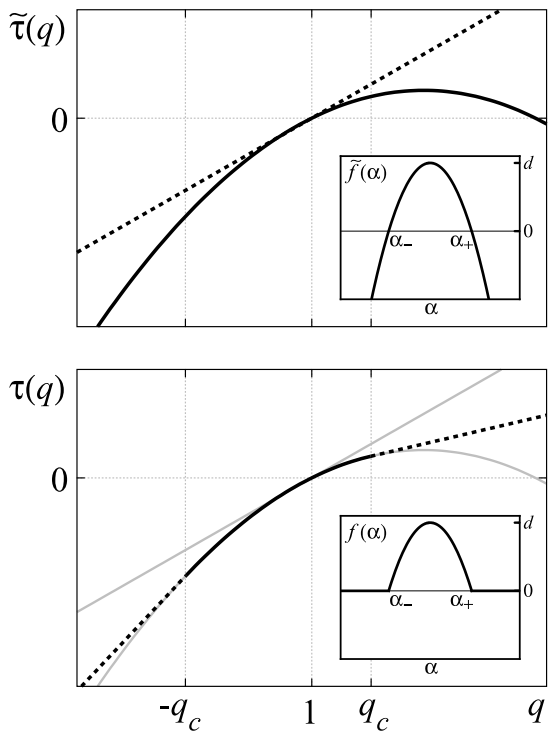}
\caption{
Sketch of the multifractal spectra at the unitary class Anderson MIT.
In the top panel, the heavy solid line corresponds to the average spectrum 
$\tilde{\tau}(q)$, defined by Eq.~(\ref{tauavg}) in the text,
as obtained at the lowest non-trivial order in the $\epsilon$-expansion 
[Eqs.~(\ref{AvgMFCSpecI})--(\ref{XiDef})].\cite{Pruisken85,HofWegner86,Wegner87}
For comparison, the heavy dashed line in the same plot is the linear spectrum for a 
plane wave state, $\tilde{\tau}(q) = \tau(q) = d(q-1)$.
In the bottom panel, the solid and dashed heavy line segments represent the typical 
spectrum $\tau(q)$, defined by Eq.~(\ref{tautyp}), as obtained in this paper via
the functional renormalization group [see Eq.~(\ref{tau(q)typFinalC})].
For $q_{c}^{-} \leq q \leq q_{c}^{+}$
(solid segment of the curve in the bottom panel), 
the typical and 
average spectra coincide. By contrast, 
the typical spectrum is linear for $q < q_{c}^{-}$, $q> q_{c}^{+}$ 
(beyond ``termination''), 
as depicted by the dashed curve segments in the bottom panel. 
The two curves in the top panel are rendered as faint gray lines in the bottom, for comparison.
The inset in the top (bottom) panel depicts the average (typical) singularity spectrum
at the unitary class MIT corresponding to the as-sketched $\tilde{\tau}(q)$ [$\tau(q)$].
\label{FigTauq}}
\end{figure}

For 
$ q > q_c^{+}$, $ q < q_c^{-}$,
the typical spectrum
$\tau(q)$ deviates completely from the average
$\tilde{\tau}(q)$, given 
by  Eqs.~(\ref{AvgMFCSpecI})--(\ref{XiDef})
to lowest order in the epsilon expansion.
Indeed, it can be rigorously proved\cite{Janssen94} that the $\tau(q)$ spectrum (as defined 
for a typical wavefunction) must be a monotonically increasing function of $q$; by comparison, 
the average spectrum $\tilde{\tau}(q)$ in 
Eqs.~(\ref{AvgMFCSpecI})--(\ref{XiDef})
is monotonically \emph{decreasing} for $q > (d + \Xi)/2 \Xi$.
For 
$ q > q_c^{+}$, $ q < q_c^{-}$,
the rare maxima (minima)
of the wavefunction 
amplitude
dominate the IPR 
[Eq.~(\ref{IPRDef})], as computed for a representative wavefunction in a 
fixed disorder realization. In this regime, the associated $\tau(q)$ is \emph{linear} in $q$.
By contrast, Eq.~(\ref{TypISAvg}) holds for $q_c^{-} < q < q_c^{+}$.
In Fig.~\ref{FigTauq}, we plot the average spectrum $\tilde{\tau}(q)$ as given by
Eqs.~(\ref{AvgMFCSpecI})--(\ref{XiDef}), 
as well as our final result for the typical spectrum $\tau(q)$, 
which we obtain in Sec.~\ref{sec: FRG} of this paper 
[Eq.~(\ref{tau(q)typFinalC}), below].

We say that the multifractal behavior of the typical $\tau(q)$ spectrum ``terminates'' 
at $q=q_c^{\pm}$.
This result in turn implies that the singularity spectrum $f(\alpha)$ must 
also suffer ``termination,'' i.e., vanish outside of the range bounded by $\alpha_{\pm}$. 
The paramount distinction between typical vs.\ average spectra is therefore summarized as follows:
the termination of $\tau(q)$ and $f(\alpha)$ reflects the dominance of \emph{rare 
amplitude
extrema} occuring in a representative wavefunction computed for a particular configuration of the disorder, 
whereas the deviation of $\tilde{\tau}(q)$ and $\tilde{f}(\alpha)$ from the former reflects the influence 
of \emph{rare disorder realizations} that enter into the averaged IPR, $\overline{P_{q}}$
[Eq.~(\ref{tauavg})].

\subsection{Operator product expansion and the functional renormalization group\label{IntroOPE}}

The dimension $x^*_q$ ($q \in \{1,2,\ldots\}$) in Eqs.~(\ref{AvgMFCSpecI}) and 
(\ref{AvgMFCSpecxq})
describes the scaling of 
the disorder-averaged $q^{\mathrm{th}}$ LDOS moment at criticality,
represented by the operator $\mathcal{O}_{q}(\vex{r})$.
In order to extract the evolution of the typical value of an LDOS moment,
we require a scaling equation for its entire probability distribution:
a functional RG approach
will turn out to be necessary.
We will demonstrate that the scaling of the typical LDOS moments 
determines the $\tau(q)$ spectrum.

A key technical difference
distinguishing
the calculation of the typical $\tau(q)$ spectrum
from that of the average $\tilde{\tau}(q)$ spectrum 
is that 
different 
LDOS
moments couple to each other along the 
FRG flow.
This coupling among the moments is encoded in the operator 
product expansion (OPE) of the scaling operators
at the delocalization critical point,
\begin{equation}
	\mathcal{O}_{q}(\vex{r})
	\,
	\mathcal{O}_{q'}(\vex{r'})
	=
	\frac{C_{q, q'}^{q+q'}}
	{|\vex{r}-\vex{r'}|^{x^{*}_{q}+x^{*}_{q'}-x^{*}_{q+q'}} }
	\mathcal{O}_{q+q'}\left(\frac{\vex{r}+\vex{r'}}{2}\right)
	+\ldots
	\label{OPEDef}
\end{equation}
Whenever the OPE coefficient $C_{q, q'}^{q+q'}\neq 0$, lower moments 
generate
higher ones upon the RG transformation.
The $\{x^*_q\}$ satisfy the convexity relation\cite{Duplantier91}
\begin{equation}\label{ConvexityIntro}
	x^*_{q+q'} < x^*_{q}+x^*_{q'} < 0  
\end{equation}
for $q, q' > 1$.
Since the $\{x^*_q\}$ are negative here, Eq.~(\ref{ConvexityIntro}) indicates that
higher moments are much more relevant,
and hence we are forced to retain all mutually coupled moments in the theory,
without being able to resort to truncation.
The FRG allows us to organize and track the entire infinite tower of 
LDOS moment operators.
The non-zero OPE coefficient $C_{q, q'}^{q+q'}$ 
leads to a non-linearity within the FRG;
the unbounded broadening suggested by the $q$-dependence of $x^*_q$
(reflecting the ever more relevant nature of the corresponding operators,
with increasing $q$) 
is balanced by this non-linearity. 
For small enough values of $q$, the non-linearity will entirely offset
the unbounded broadening and render it  inconsequential,
whereas for sufficiently large values of $q$
this will result in the termination of the typical $\tau(q)$ spectrum.

The mechanism described above is known to be responsible for the termination
of the multifractal spectrum in a 
special (so-called `chiral')\cite{DisClasses} 
symmetry class of 2D models, possessing quenched disorder.
Carpentier and Le Doussal\cite{Carpentier00,Carpentier01} pioneered 
the use of the FRG technique in their study of the random phase XY (gauge glass) model.
This 
method
was later applied\cite{Carpentier01} to the
problem\cite{Ludwig94} of a 2D massless Dirac fermion, 
subject to a static, random Abelian vector potential.
The FRG provided direct confirmation of the multifractal termination
for this problem,
a result previously conjectured\cite{Chamon96,Castillo97} for the vector potential model.
Later, Mudry et al.\cite{Mudry03} extended the FRG to
a more general 2D disordered Dirac model belonging
to the symmetry class $BD$I (chiral orthogonal symmetry class).
[We have adopted the nomenclature for quantum disorder classes employed
in Ref.~\onlinecite{DisClasses}.]
In these works, the FRG equation  
constructed from the set of operator scaling dimensions
$x_{q}^{*}$ and OPE coefficients $C_{q,q'}^{q+q'}$ [Eq.~(\ref{OPEDef})] 
takes the form of 
the so-called  Kolmogorov-Petrovsky-Piscounov (KPP) equation,\cite{RefKPP}
which describes non-linear diffusion
in one dimension.
It is the non-trivial behavior of the
long-time asymptotics of the
solution to the KPP equation 
that is responsible for the termination. We will show that the same
equation arises in the 
general
context of the
typical $\tau(q)$ spectrum in
the unitary 
symmetry class, 
at the Anderson MIT critical point in $d = 2 + \epsilon$
(with obvious extensions to additional symmetry classes).

\subsection{Outline}

Using the framework of the fermionic replica (compact)
NL$\sigma$M approach,\cite{Wegner79,EfetovLarkinKhmelnitskii80,LeeRamakrishnan}
we compute the OPE coefficient $C_{q,q'}^{q+q'}$ at
the critical point in $d = 2 + \epsilon$ for the unitary class. 
Combining this result with the scaling dimensions given by 
Eqs.~(\ref{AvgMFCSpecxq})--(\ref{XiDef}),
we formulate the functional 
renormalization group for the tower of LDOS moment operators
that enters into the computation of the
typical $\tau(q)$ spectrum.  
Then we
use the FRG to demonstrate that the same
mechanism 
active in the 2D Dirac models,\cite{Carpentier00,Carpentier01,Mudry03} 
discussed above, leads to the termination of the multifractal spectrum
at the MIT.
We obtain the $\tau(q)$ spectrum for a typical
wavefunction, which agrees with previous heuristic
arguments.\cite{MirlinEvers00,EversMirlin00}

The rest of the paper is organized as follows:

In Sec.~\ref{sec: definitions and model}, we review the connection between the IPR and 
the local density of states, and we introduce a generating function that will be used to determine 
the typical $\tau(q)$ spectrum. We then establish conventions for the 
fermionic replica NL$\sigma$M, and identify the composite operators that represent moments of the
local density of states in the low-energy field theory. 
In Sec.~\ref{sec: FRG} we use the 
operator product expansion (OPE)
of the LDOS moment operators at the MIT as input
into the FRG, which then allows us to compute the scaling behavior of the generating 
function introduced in Sec.~\ref{sec: definitions and model}. We thereby obtain
the typical $\tau(q)$ spectrum. We discuss our results and draw conclusions in 
Sec.~\ref{sec: conclusion}. 

The derivation of the OPE of the operators $\{{\mathcal O}_q({\bf r})\}$
representing the LDOS moments, 
which constitutes the technical field 
theoretic content of this work, has been relegated to 
Sec.\ \ref{sec: model renomaliation and composite operator renomalization}.
In this Section, we rederive the anomalous scaling dimensions of the LDOS moment operators,
and we compute the required OPE coefficient between properly normalized 
versions of these. The results obtained are invoked as needed in the earlier Sec.~\ref{sec: FRG}, 
so the reader less interested in calculational details may skip 
Sec.\ \ref{sec: model renomaliation and composite operator renomalization} entirely.


\section{Definitions and Model \label{sec: definitions and model}}

\subsection{
Extracting multifractality from the LDOS -- typical spectra
\label{sec: IPRtoLDOS}}

Consider the local density of states (LDOS), defined as
\begin{align}\label{LDOSDef}
	\nu(\varepsilon,\vex{r}) 
	&=
	\frac{-1}{\pi} \mathrm{Im} \, G_{R}(\varepsilon;\vex{r},\vex{r})
	\nonumber\\
	&=
	\sum_{i}
	\delta(\varepsilon - \varepsilon_{i})
	|\psi_{i}(\vex{r})|^{2},
\end{align}
where the retarded Green's function is given by
\begin{equation}\label{GRDef}
	G_{R}(\varepsilon;\vex{r},\vex{r'})
	=
	\sum_{i}
	\frac{\psi_{i}(\vex{r}) \psi^{*}_{i}(\vex{r'})}
	{\varepsilon - \varepsilon_{i} + i \eta},
\end{equation}
with $\eta \rightarrow 0^{+}$. 

On the metallic side of the delocalization transition, we cannot relate 
$P_{q}(\varepsilon)$, defined in terms of a single wavefunction by Eq.~(\ref{IPRDef}),
directly to the LDOS.\cite{Wegner80} In order to use the field theory
approach, we require that the LDOS constitute a smooth, well-defined function of energy in a 
closed, finite-size system; this necessitates the 
retention of the finite energy level 
broadening
$\eta \gtrsim \Delta$, 
where $\Delta$ is the global level spacing. 
(Although a formal device in this context, the 
broadening may
be attributed to, e.g.,
inelastic relaxation processes neglected in the non-interacting, 
single particle approach.)

We define\cite{Wegner80}
\begin{align}\label{IPRtoLDOS}
	\frac{1}{L^{d(q - 1)} \, p^{(q)}(\varepsilon)}
	\equiv
	\frac{\int \dvex{r} \, \nu^{q}(\varepsilon,\vex{r})} 
	{\left[\int \dvex{r} \, \nu(\varepsilon,\vex{r})\right]^{q}}.
\end{align} 
The quantity $p^{(q)}(\varepsilon)$ denotes the participation ratio, which
receives contributions from states with energies residing in a window
of width 
$\eta$
about $\varepsilon$. On the metallic side of the transition,
the right-hand side (RHS) of Eq.~(\ref{IPRtoLDOS}) should scale identically as 
Eq.~(\ref{IPRDef}).\cite{Wegner80} 

It will prove useful to introduce the moment generating function
for the $q^{\mathsf{th}}$ power of the LDOS,
\begin{equation}\label{FGenDef}
	F_{q}(\xi; L) \equiv 
	\left\langle 
	\exp\left[ 
	- \xi \int \dvex{r} \, \nu^{q}(\varepsilon,\vex{r})
	\right]
	\right\rangle,
\end{equation}
where the angle brackets $\langle \cdots \rangle$ denote a suitable ensemble average
over realizations of the quenched disorder; $L$ is the linear system size. 
Using the identity
\begin{equation}\label{LogIntIdent}
	\ln \phi = \int_{0}^{\infty} \frac{d \xi}{\xi} 
	\left(
	e^{-\xi} - e^{-\phi \, \xi}
	\right)
\end{equation}
and replacing 
$P_{q}$ with the RHS of Eq.~(\ref{IPRtoLDOS})
in Eq.~(\ref{tautyp}), the typical multifractal spectrum exponent $\tau(q)$
may be written as
\begin{equation}\label{tau(q)FromFGen}
	\tau(q) = 
	\frac{d\phantom{\ln}}{d \ln L}
	\int_{0}^{\infty} \frac{d \xi}{\xi} 
	\left[
	F_{q}(\xi;L) - q\,F_{1}(\xi;L)
	\right].
\end{equation}
Our goal is to compute the scaling behavior of 
the moment generating function $F_{q}(\xi;L)$, 
and thereby obtain 
the typical $\tau(q)$ spectrum via 
Eq.~(\ref{tau(q)FromFGen}).
In closing this subsection, we note that the evaluation of Eq.~(\ref{tau(q)FromFGen}) using 
the lowest order cumulant expansion for $F_{q}(\xi; L)$ recovers the 
average $\tilde{\tau}(q)$ spectrum, 
Eq.~(\ref{AvgMFCSpecI}); we will discuss this point in detail in Sec.~\ref{sec: FRG}.

\subsection{NL$\sigma$M formulation \label{sec: definitions of NLSM}}

We examine in this paper the properties of the multifractal spectrum at the 
Anderson MIT in the unitary 
symmetry
class. The critical point itself is accessed 
via the standard perturbative $\epsilon$-expansion in $d = 2 + \epsilon$ dimensions,
with $0 < \epsilon \ll 1$.
Our low-energy, effective field theory starting point is the \emph{compact} replica 
NL$\sigma$M,\cite{Wegner79,EfetovLarkinKhmelnitskii80,LeeRamakrishnan} 
defined
by the functional integral 
\begin{align}
	Z
	& \equiv \int \mathcal{D}[\Qh] e^{-S},
	\nonumber\\
\intertext{where}
	S[\Qh] 
	& \equiv
	\frac{1}{2t}
	\int \dvex{r}
	\Tr
	\left(
	\bm{\nabla}\Qh\cdot\bm{\nabla}\Qh
	\right)
	-
	h
	\int \dvex{r} 
	\Tr
	\left(
	\Lambdah_z \Qh 
	\right).\label{SDef}
\end{align}
In this equation, the 
``temperature'' $t$ is inversely
proportional to the dimensionless dc conductance of the disordered metal, while the 
``external field'' $h$ 
serves as an infrared regulator,
coupling to the local density of states (LDOS) operator, as defined below.
The symbol $\Qh$ denotes a $2n\times 2n$ Hermitian matrix field satisfying
\begin{equation}\label{QConstraint}
	\Qh^{2}(\vex{r})=\hat{\mathbb{I}}_{2n},
	\quad 
	\Tr \Qh(\vex{r}) = 0.
\end{equation}
The constant matrix 
\begin{equation}\label{LambdazDef}
	\Lambdah_z = 
	\mathrm{diag}\,\left( \hat{\mathbb{I}}_n,-\hat{\mathbb{I}}_n \right)
\end{equation}
sets the (trivial) saddle-point for the action defined by Eq.~(\ref{SDef}).
The identity in the space of $2n\times 2n$ and $n\times n$ square matrices
is denoted by $\hat{\mathbb{I}}_{2n}$ and $\hat{\mathbb{I}}_{n}$ in Eqs.~(\ref{QConstraint}) 
and (\ref{LambdazDef}), respectively.
In these equations,
$n$ is proportional to
the number of replicas, with $n \rightarrow 0$ at the end of the calculation.\cite{Wegner79,LeeRamakrishnan}
The target space of the NL$\sigma$M is the 
compact
coset $\mathrm{G}(2 n)/\mathrm{G}(n)\times \mathrm{G}(n)$,
where $\mathrm{G}=\mathrm{O},\mathrm{U},\mathrm{Sp}$ 
for the orthogonal, unitary, and symplectic symmetry classes, respectively. 
In the following, we focus upon the unitary universality class, $\mathrm{G}=\mathrm{U}$.
The field theory in Eqs.~(\ref{SDef}) and (\ref{QConstraint}) can be derived\cite{EfetovLarkinKhmelnitskii80} 
from a microscopic Grassmann path integral describing a system of non-interacting fermions, lacking
time-reversal invariance, averaged over configurations of a Gaussian, white noise-correlated random
potential.

We employ `$\sigma$-$\pi$' coordinates\cite{McKaneStone81} on the target manifold,
\begin{equation}\label{WDef}
	\Qh =
	\left[
	\begin{array}{cc}
	\left(\hat{\mathbb{I}}_{n} - \Wh \Wh^{\dag} \right)^{1/2} & \Wh \\
	\Wh^{\dag} & -\left(\hat{\mathbb{I}}_{n} - \Wh^{\dag} \Wh \right)^{1/2}
	\end{array}
	\right].
\end{equation}
For the unitary class, $\Wh(\vex{r})\rightarrow W^{\alpha}{}_{\beta}(\vex{r})$ is an unconstrained, 
complex-valued matrix, with $\alpha,\beta \in \{1,\ldots,n\}$.

Non-interacting electrons residing in $d > 2$ spatial dimensions and subject to quenched
disorder possess a diffusive metallic phase, defined as the presence of extended wavefunctions
at the Fermi energy, provided that the disorder is sufficiently weak. The disorder strength 
is quantified by the ``bare'' conductance at the scale of the mean free path, proportional 
to $1/t$ in the effective field theory [Eq.~(\ref{SDef})]. In direct analogy with the 
$\mathrm{O}(3)/\mathrm{O}(2)$ NL$\sigma$M description of classical magnetic 
ordering,\cite{Polyakov75,BrezinZinnJustinLeGuillou76,FieldTheory} the 
``low temperature'' (weak disorder) regime $0 \leq t < t^{*}$ of the model in Eq.~(\ref{SDef}) exhibits 
spontaneous continuous symmetry breaking, so that the `$\sigma$' fields $(\hat{\mathbb{I}}_{n} - \Wh \Wh^{\dag})^{1/2}$
and $(\hat{\mathbb{I}}_{n} - \Wh^{\dag} \Wh )^{1/2}$, which form the diagonal elements of the $\Qh$-matrix
in the parameterization of Eq.~(\ref{WDef}), acquire non-zero expectation values throughout the diffusive 
metallic phase. By contrast, the off-diagonal `$\pi$' fields $\Wh$ and $\Wh^{\dagger}$ represent 
small spatial fluctuations with vanishing mean in this regime. 
Here, $t = t^{*}>0$ locates the MIT in $d = 2 + \epsilon$.

An unusual aspect of the theory of the MIT transcribed in Eq.~(\ref{SDef}) is the fact that this 
spontaneous symmetry breaking occurs also at the delocalization transition itself ($t = t^{*}$), and survives
even into the insulating (``high temperature'') phase ($t > t^{*}$).\cite{Wegner81,McKaneStone81}
In the effective NL$\sigma$M field theory, the trace of the matrix $\Lambdah_{z} \Qh(\vex{r})$ represents 
the LDOS $\nu(\varepsilon,\vex{r})$ [Eq.~(\ref{LDOSDef})] for the disordered electron system:\cite{Wegner80}
\begin{align}\label{LDOSOpDef}
	\nu(\varepsilon,\vex{r}) 
	&\sim
	\Tr\left[\Lambdah_z \Qh(\vex{r}) \right]
	\nonumber\\
	&=
	\Tr\left\{\!
	\left[\hat{\mathbb{I}}_{n} - \Wh \Wh^{\dag}(\vex{r}) \right]^{1/2}
	+
	\left[\hat{\mathbb{I}}_{n} - \Wh^{\dag} \Wh(\vex{r}) \right]^{1/2}
	\right\}.
\end{align}
This is the same operator that appears in the action Eq.~(\ref{SDef}), where it couples to the external 
field parameter $h$. While the \emph{character} of the typical wavefunction changes from extended to localized
upon traversing the mobility edge, as encoded by, e.g., the typical multifractal exponent $\tau(q)$ for $q \geq 2$
[Eq.~(\ref{tau(q)FromFGen})],
the average density of states does not exhibit critical behavior across the transition.\cite{Wegner81}
The LDOS operator on the RHS of Eq.~(\ref{LDOSOpDef}) retains a non-zero expectation value so long as
the average density of states is non-vanishing; consequently, the $\Qh$-matrix cannot be interpreted as an order 
parameter for the MIT. Technically, this result (an exception to Goldstone's theorem)\cite{McKaneStone81}
obtains from the NL$\sigma$M only \emph{after} the replica limit $n \rightarrow 0$ is taken.

For any non-zero, integral number of replicas $n \in \{1,2,\ldots\}$, the model in Eq.~(\ref{SDef}) also
possesses a (different) second order transition at $t = t_{n}^{*} > 0$, separating a low temperature
``ferromagnetic'' phase ($t < t_{n}^{*}$) from the high temperature ``paramagnet'' ($t > t_{n}^{*}$). 
In contrast to the replica limit $n \rightarrow 0$ appropriate to the description of electronic wavefunction 
(de)localization, the NL$\sigma$M with $n \geq 1$ is characterized by a \emph{restoration} of the symmetry at 
the critical point between the `$\sigma$' (diagonal) and `$\pi$' (off-diagonal) components of the $\Qh$-matrix, 
within the parameterization given by Eq.~(\ref{WDef}). This is the conventional behavior expected for a classical 
statistical mechanics model describing spontaneous continuous symmetry breaking in the vicinity of the critical 
point. 

Let us assume that we are interested only in properties of the NL$\sigma$M given by Eq.~(\ref{SDef}) at the 
critical point, $t = t_{n}^{*}$. Because the symmetry is restored at the transition, for non-zero $n$ we are 
permitted to make the following $\mathrm{U}(2 n)$ ``rotation'' from $\Lambdah_z$ to $\Lambdah_{x}$ in 
Eq.~(\ref{LDOSOpDef}):
\begin{align}
	\nu(\varepsilon,\vex{r}) 
	\sim
	\Tr
	\left[\Lambdah_z \Qh(\vex{r})\right]
	& \to
	\Tr
	\left[\Lambdah_x \Qh(\vex{r})\right]
	\nonumber\\
	& =
	\Tr
	\left[\Wh(\vex{r}) + \Wh^{\dagger}(\vex{r})\right],
	\label{eq: LDOS rotated}
\end{align}
where $\Lambdah_{x}$ denotes the block Pauli matrix generalizing 
Eq.~(\ref{LambdazDef}), in the standard basis.

In the technical field theoretic portion of this paper, 
Sec.~\ref{sec: model renomaliation and composite operator renomalization}, 
we employ the NL$\sigma$M defined by Eqs.~(\ref{SDef})--(\ref{WDef}) to extract the properties of the LDOS 
operator and its moments. 
Our strategy is to work, as usual,  at fixed,
integral $n \geq 1$ throughout the intermediate 
stages of our computations. At the critical point in $d = 2 + \epsilon$, we are then free to employ the LDOS 
representation given by the RHS of Eq.~(\ref{eq: LDOS rotated}). Only at the end of our work will we perform 
the required analytic continuation $n \rightarrow 0$ (which smoothly deforms $t_n^{*} \rightarrow t^{*}$), so 
as to obtain
(perturbative)
results appropriate to the MIT.

\subsection{LDOS moments as composite eigenoperators \label{sec: tildetau spectrum}}

Higher integral moments of the LDOS can be similarly represented by local composite 
operators in the NL$\sigma$M.
The renormalization group (RG) transformation does not preserve the form of an operator 
\begin{equation}\label{nuMoment}
	\nu^p = \left[\Tr\left(\Wh + \Wh^\dagger\right)\right]^p,
\end{equation}
obtained by taking a 
power
of Eq.\ (\ref{eq: LDOS rotated}).
Nevertheless, such a structure can be decomposed into invariant eigenoperators, 
each of 
which possessing an independent scaling dimension.

This idea is most easily understood via analogy to the simpler $\mathrm{O}(3)/\mathrm{O(2)}$
model,\cite{Polyakov75,BrezinZinnJustinLeGuillou76,FieldTheory}
to which the field theory defined by Eqs.~(\ref{SDef})--(\ref{WDef}) reduces for the case of $n = 1$ 
[since $\mathrm{U}(2)/\mathrm{U}(1)\times\mathrm{U}(1) \sim \mathrm{SU}(2)/\mathrm{U}(1) 
\sim \mathrm{O}(3)/\mathrm{O}(2)$].
In this NL$\sigma$M, the target manifold is simply the two-sphere, parameterized by
the unconstrained transverse coordinates $\pi_{\pm} \equiv \pi_{x} \pm i \pi_{y}$, 
with $z$-component $\sigma = \sqrt{1 - \pi_{+} \pi_{-}}$. A complete basis of
local eigenoperators with no derivatives is the set of ordinary spherical 
harmonics $\{Y_{l,m}(\pi_{+},\pi_{-},\sigma)\}$. 
All operators belonging to a given irreducible representation of the symmetry group 
possess
the same renormalization; 
therefore, any linear combination of spherical harmonics sharing a common $l$
value constitutes an eigenoperator. 
The field coordinates $\pi_{\pm}$ are themselves 
eigenoperators belonging to $l = 1$, as is the combination
\begin{equation}
	\nu \equiv \pi_{+} + \pi_{-} \propto Y_{1,-1} - Y_{1,1}.
\end{equation}
For an arbitrary integer moment of $\nu$, one can use angular momentum addition to
establish the decomposition
\begin{equation}\label{O(3)EigenOpDecomp}
	(\pi_{+} + \pi_{-})^l = \sum_{j = 0}^{l} \mathcal{O}_{j}^{(l)}, 
\end{equation}
where the eigenoperators $\mathcal{O}_{j}^{(l)}$ are defined via
\begin{equation}\label{O(3)EigenOpA}
	\mathcal{O}_{j}^{(l)} 
	= 
	\sum_{m = - j}^{j} \kappa_{j,m}^{(l)}  \, Y_{j,m}(\pi_{+},\pi_{-},\sigma).
\end{equation}
with certain coefficients $ \kappa_{j,m}^{(l)}$.
For the highest total angular momentum block $j = l$ in Eq.~(\ref{O(3)EigenOpDecomp}),
one has
\begin{equation}\label{O(3)EigenOpB}
	\mathcal{O}_{l}^{(l)} 
	=
	\left(\pi_{-}^{l} + \ldots + \pi_{+}^{l} \right),	
\end{equation}
since the ``highest and lowest weight states'' $\pi_{+}^l$ and $\pi_{-}^l$
are eigenoperators proportional to $Y_{l,l}$ and $Y_{l,-l}$, respectively.

The coefficients $\{\kappa_{j,m}^{(l)}\}$ on the RHS of Eq.~(\ref{O(3)EigenOpA}) are
determined entirely by group theory (i.e., are composed of sums of
products of appropriate Clebsch-Gordan coefficients),\cite{footnote-Clebsch}
up to an overall
$m$-independent normalization for all operators belonging to a given 
total angular momentum block $j$. This normalization can be established via the
convention
\begin{equation}\label{lambdaNormDef}
	Y_{l,-l} \equiv \lambda_{l} \, \pi_{-}^{l}.
\end{equation} 
In a similar fashion, the operator
in Eq.~(\ref{nuMoment}) should be decomposed into a sum of terms belonging
to different irreducible representations of the group $\mathrm{U}(2 n)$. Each
such term can be further decomposed into a linear combination of basis
operators with appropriate ``magnetic'' quantum numbers 
determined by the transformation properties under the 
subgroup $\mathrm{U}(n)\times\mathrm{U}(n)$.

It is useful to push this analogy a little further. 
In order to extract the typical $\tau(q)$ spectrum in the unitary class model,
we need the scaling dimension of the most relevant eigenoperator (in the RG sense) 
contributing to each of the $p^{\mathrm{th}}$ LDOS moments in Eq.~(\ref{nuMoment}),
$p \in \{1,2,\ldots\}$, as well as the operator product expansion (OPE) between pairs of such most
relevant eigenoperators. The most relevant eigenoperator contributing to the 
decomposition of Eq.~(\ref{nuMoment}), for a given fixed $p$, is analogous to the highest 
(total) angular momentum operator $\mathcal{O}^{(l)}_{l}$ contributing to the $l^{\mathrm{th}}$ 
moment of $(\pi_{+} + \pi_{-})$ in Eq.~(\ref{O(3)EigenOpDecomp}),\cite{footnote-O(3)vsAnderson} with $l = p$.
[Precise definitions of the eigenoperators that we employ in the 
$\mathrm{U}(2n)/\mathrm{U}(n)\times\mathrm{U}(n)$ NL$\sigma$M are given by Eqs.~(\ref{LDOSEigenOps}) 
and (\ref{LDOSEigenOpsSummed}), below.]
In the $\mathrm{O}(3)/\mathrm{O}(2)$ model, we can effectively trade the operator $\mathcal{O}^{(l)}_{l}$,
which for large $l$ is a complicated sum of many terms according to Eq.~(\ref{O(3)EigenOpA}),
for its lone ``lowest weight state'' component $Y_{l,-l}$ [Eqs.~(\ref{O(3)EigenOpB}) and (\ref{lambdaNormDef})]. 
Obviously, both operators share the same scaling dimension. Moreover, the structure of the
OPE between $\mathcal{O}^{(l)}_{l}$ and $\mathcal{O}^{(l')}_{l'}$ follows from that
of the product between their lowest weight state constituents. 
Consider the following OPE \emph{at zero coupling} ($t = 0$):
\begin{align}\label{OPEZeroCoupling}
	\lambda_{l} \lambda_{l'} \mathcal{O}_{l}^{(l)} \mathcal{O}_{l'}^{(l')} &= c_{l,l'}^{l+l'} \lambda_{l+l'} \mathcal{O}_{l+l'}^{(l+l')},
	\nonumber\\
	\left(Y_{l,-l} + \ldots \, \right) \left(Y_{l',-l'} + \ldots \, \right) &= c_{l,l'}^{l+l'} \left(Y_{l+l',-l-l'} + \ldots \, \right),
\end{align}
where we have defined the OPE coefficient 
\begin{equation}\label{OPECoeffZeroCoupling}
	c_{l,l'}^{l+l'} \equiv \frac{\lambda_{l} \lambda_{l'}}{\lambda_{l+l'}}.
\end{equation}
The crucial point is that the relative weight of each term appearing in the expansion for the
eigenoperator $\mathcal{O}_{l}^{(l)}$ [Eq.~(\ref{O(3)EigenOpA})] is entirely fixed by group theory; only the 
overall, $l$-dependent normalization is arbitrary. The required OPE coefficient in Eq.~(\ref{OPECoeffZeroCoupling})
is then determined by just this normalization for the lowest weight state operators, Eq.~(\ref{lambdaNormDef}). 
Of course, 
this argument neglects loop corrections, 
which may modify
the value of the OPE coefficient 
given by Eq.~(\ref{OPECoeffZeroCoupling}),
computable systematically within the $\epsilon$-expansion.
This, however, 
cannot alter the structure of 
Eq.~(\ref{OPEZeroCoupling}).

With the above in mind, we consider the component
\begin{equation}\label{WMoment}
	\left[\Tr \Wh \right]^p
\end{equation}
of the LDOS moment in Eq.~(\ref{nuMoment}). 
As opposed to
the sphere model discussed above,
this pure $\Wh$
power does not represent an eigenoperator for $n > 1$. However, a useful
subset\cite{footnote-OtherEigenOps} of the RG eigenoperators \emph{can} be built out of
$p$-fold products of `$\pi$' ($W^{\alpha}{}_{\beta}$) field matrix elements:
\begin{align}\label{LDOSEigenOpDef}
	\mathcal{O}^{\; \alpha_{1} \alpha_{2} \ldots \alpha_{p}}_{p \; (\beta_{1} \beta_{2} \ldots \beta_{p})_\mathcal{Y}}(\vex{r})
	\equiv
	\frac{1}{p!}
	W^{\alpha_1}{ }_{(\beta_1} 
	W^{\alpha_2}{ }_{\beta_2} 
	\cdots 
	W^{\alpha_p}{ }_{\beta_p )_\mathcal{Y}},
\end{align}
where $(\cdots )_{\mathcal{Y}}$ means a suitable symmetrization prescribed by a Young tableau $\mathcal{Y}$.
For fixed $p$, the most relevant operator (in the sense of the RG, at the MIT in $d = 2 + \epsilon$) 
is given by the totally antisymmetric Young tableau,\cite{Wegner80,Pruisken85}
\begin{align}\label{LDOSEigenOps}
	&\mathcal{O}^{\; \alpha_{1} \alpha_{2} \ldots \alpha_{p}}_{p \; [\beta_{1} \beta_{2} \ldots \beta_{p}]}(\vex{r})
	\nonumber\\
	&\;\;\,\equiv
	\left(\frac{1}{p!}\right)^{2}
	\sum_{\bm{\mathrm{P}}} 
	\sgn(\bm{\mathrm{P}})\,
	\left[
	W^{\alpha_1}{ }_{\beta_{\bm{\mathrm{P}}(1)}} 
	\cdots 
	W^{\alpha_p}{ }_{\beta_{\bm{\mathrm{P}}(p)}}
	\right],
\end{align}
with $\bm{\mathrm{P}}$ a permutation of $p$ symbols;
$\sgn(\bm{\mathrm{P}})$ denotes the sign of the permutation.
Because of the antisymmetrization requirement, each distinct operator defined through
Eq.~(\ref{LDOSEigenOps}) is identified by any permutation of a complete set of indices 
$\{\alpha_{i}\}$ satisfying $\alpha_{1}\neq\alpha_{2}\neq\ldots\neq\alpha_{p}$, 
and similarly for the $\{\beta_{i}\}$. Indices range from $1$ to $n$, so that many 
different operators can be associated to each integral moment of the LDOS, at 
least for sufficiently large $n$.

Physically, we would like establish a one-to-one correspondence between the $p^{\mathrm{th}}$ 
LDOS moment $[\nu(\varepsilon,\vex{r})]^{p}$ in the disordered electron system, 
and a single, \emph{unique} operator $\mathcal{O}_{p}$ in the NL$\sigma$M field theory that 
represents its most relevant component. This can be accomplished 
by tracing over pairs 
of indices in Eq.~(\ref{LDOSEigenOps}) in the following fashion:
\begin{align}\label{LDOSEigenOpsSummed}
	\mathcal{O}_{p}(\vex{r}) 
	&\equiv 
	\sum_{\alpha_{1} = 1}^{n}\ldots\sum_{\alpha_{p} = 1}^{n}
	\mathcal{O}^{\; \alpha_{1} \alpha_{2} \ldots \alpha_{p}}_{p \; [\alpha_{1} \alpha_{2} \ldots \alpha_{p}]}(\vex{r}).
\end{align}
With this definition, the eigenoperators
\begin{gather}
	\mathcal{O}_{2} 
	= 
	\frac{
	\left[\Tr\left(\Wh \right)\right]^2 - \Tr\left(\Wh^2 \right)
	}
	{(2!)^2},
	\nonumber\\
	\mathcal{O}_{3} 
	= 
	\frac{
	\left[\Tr\left(\Wh \right)\right]^3 
	- 3 \Tr\left(\Wh^2 \right) \Tr\left(\Wh \right)
	+ 2 \Tr\left(\Wh^3 \right)
	}
	{(3!)^2},
	\nonumber
\end{gather}
etc., are easily recognized as natural deformations of the LDOS moments obtained by
taking powers of Eq.~(\ref{WMoment}).\cite{footnote-OtherEigenOps}
Moreover, we will establish in 
Sec.~\ref{sec: model renomaliation and composite operator renomalization}
that the set $\{\mathcal{O}_{p}\}$ closes under the operator product expansion (OPE), 
up to less relevant operators generated on the right-hand side of Eq.~(\ref{OPEDef}), 
which we may ignore. 
This is a sufficient condition to apply the functional renormalization group method.

In summary, the operators defined by Eqs.~(\ref{LDOSEigenOps}) or (\ref{LDOSEigenOpsSummed}) 
constitute the most relevant component(s) of the $p^{\textrm{th}}$ moment of the 
LDOS\cite{Wegner80,Pruisken85,footnote-OtherEigenOps}
at the MIT, and hence dominate its scaling behavior there.

\subsection{Augmented NL$\sigma$M \label{sec: tau spectrum}}

At the 
metal-insulator critical point,
the scaling of the average IPR $\overline{P_{q}}$ 
[i.e., the multifractal exponent $\tilde{\tau}(q)$, Eq.~(\ref{tauavg})] can be extracted solely 
from the scaling dimensions $x^{*}_{p}$ of the local composite operators 
$\mathcal{O}^{\; \alpha_{1} \alpha_{2} \ldots \alpha_{p}}_{p \;
[\beta_{1} \beta_{2} \ldots \beta_{p}]}(\vex{r})$
or $\mathcal{O}_{p}(\vex{r})$, with $p \in \{1,q\}$---this is the content of Eq.~(\ref{AvgMFCSpecI}) in the Introduction.
By contrast, the 
probability distribution functions 
of the IPR and LDOS [reflected 
by the typical
multifractal exponent $\tau(q)$,
Eq.~(\ref{tautyp})]
are described by the complicated generating function $F_{q}(\xi;L)$,
introduced in Eq.~(\ref{FGenDef}).
In the low-energy theory, $F_{1}(\xi;L)$ can be represented by
the NL$\sigma$M in Eq.~(\ref{SDef}) with a bare non-zero
external field parameter $h_{0}$ given by
\begin{equation}
	h_{0} 
	=
	-\xi. 
\end{equation}
Performing a renormalization group transformation upon the NL$\sigma$M 
with $h_{0} \neq 0$ generically produces higher powers of the LDOS operator as new perturbations
to the action $S$, so that terms of the form
\begin{align}\label{LDOSPowersGen}
	\delta S =
	-Y_{m}
	\int \dvex{r}
	\left\{
	\Tr
	\left[
	\Lambdah_{z}
	\Qh(\vex{r})
	\right]
	\right\}^{m},
\end{align}
for example, will be generated. Here, 
$Y_{m}$
is a coupling constant. 
The structure in Eq.~(\ref{LDOSPowersGen}) is not invariant under
the RG; with further iterations, it will
{\bf(a)} mix with other terms sharing the same ``engineering'' dimension, and 
{\bf(b)} fuse with other terms and with itself to produce new perturbations.
Among the flood of structures that arise, we will focus
only upon the most relevant terms that determine the leading scaling behavior for the
generating function $F_{1}(\xi;L)$ of the LDOS and its moments. 

Anticipating the generation of higher moments upon renormalization, 
we should augment the action in Eq.~(\ref{SDef}) (with $h_{0} = 0$) by a term of the form
\begin{equation}\label{LDOSEigenOpsAction}
	\delta S \equiv -\sum_{p=1}^{\infty} Y_{p}
	\int \dvex{r} \,
	\mathcal{O}_{p}(\vex{r}),
\end{equation}
where the ``traced'' moment operators $\mathcal{O}_{p}$ were defined above by Eq.~(\ref{LDOSEigenOpsSummed}).

At tree level, the operators defined by Eqs.~(\ref{LDOSEigenOps}) and (\ref{LDOSEigenOpsSummed}) 
are dimensionless, so that the corresponding coupling constants $\{Y_{p}\}$ are strongly relevant 
perturbations to the NL$\sigma$M action. As discussed in Sec.~\ref{AvgTypTerm}, they prove
even more relevant at the non-trivial fixed point (perturbatively accessible Anderson MIT).
Moreover, the higher moments are more relevant compared to the lower ones 
[Eq.~(\ref{ConvexityIntro})]. The FRG approach tracks the scaling behavior of this entire tower of 
operators, and uses this data to make non-trivial predictions about 
observable statistics, such as the \emph{typical}
LDOS. Within the FRG framework, only two pieces of information are needed: first, the scaling 
dimensions of the operators in Eqs.~(\ref{LDOSEigenOps}) and (\ref{LDOSEigenOpsSummed}), 
and second, the coefficient $C_{q,q'}^{q+q'}$ for the operator product  
$\mathcal{O}_{q}\otimes\mathcal{O}_{q'}\rightarrow \mathcal{O}_{q+q'}$, as defined by the OPE 
in Eq.~(\ref{OPEDef}). All quantities are to be evaluated at the MIT in $d = 2 + \epsilon$.

We use a two-stage approach to the renormalization of the ``extended'' NL$\sigma$M
[the action Eq.\ (\ref{SDef}) supplemented with (\ref{LDOSEigenOpsAction})].
The idea is to first locate the non-trivial metal-insulator fixed point
in $d = 2 + \epsilon$, obtained via the standard $\epsilon$-expansion by renormalizing 
the theory in Eq.~(\ref{SDef}) with $h \rightarrow 0$. 
(We will use dimensional regularization.)
We then compute the OPE [Eq.~(\ref{OPEDef})] at the MIT to the lowest non-trivial order 
in $\sqrt{\epsilon}$.
Finally, we run a `one-loop' RG calculation \emph{at this non-trivial fixed point}, 
for the full model defined by Eqs.~(\ref{SDef}) and (\ref{LDOSEigenOpsAction}). 
The required one-loop functional renormalization group equation is obtained from the OPE.\cite{Cardy96}
Note that since we are interested in LDOS and IPR statistics at the Anderson metal-insulator
transition ($t = t^{*}$), 
rather than in the diffusive metallic phase 
($t < t^{*}$), 
we are required to run the FRG at this non-trivial fixed point.\cite{WilsonianRG}

In order to streamline the presentation, the above-described field theory calculations 
are relegated to the last Sec.~\ref{sec: model renomaliation and composite operator renomalization}
of this paper. The obtained results required for the functional RG are simply invoked as needed 
in the next Sec.~\ref{sec: FRG}, so that the reader less interested in calculational details 
may avoid Sec.~\ref{sec: model renomaliation and composite operator renomalization}
entirely.


\section{Functional RG for the typical $\tau(q)$ spectrum 
\label{sec: FRG}}

\subsection{From coupled RG to KPP equations}

The typical $\tau(q)$ spectrum, defined in Sec.~\ref{Intro} by Eq.~(\ref{tautyp}),
can be extracted from the generating function $F_{q}(\xi; L)$, introduced
in Eq.~(\ref{FGenDef}). The relationship is expressed
by Eq.~(\ref{tau(q)FromFGen}).
In terms of the NL$\sigma$M formulation reviewed in 
Secs.~\ref{sec: definitions of NLSM}--\ref{sec: tau spectrum}, 
$F_{q}(\xi; L)$ may be encoded as
\begin{equation}\label{FGenNLsM}
	F_{q}(\xi; L) \sim 
	\left\langle 
	\exp\left[ 
	\sum_{p=1}^{\infty} Y_{p q}
	\int \dvex{r} \,
	\NoOp{
	\mathcal{O}_{p q}
	}
	(\vex{r})
	\right]
	\right\rangle,
\end{equation}
where $q =1,2,3,\cdots$, and $\NoOp{\mathcal{O}_{p q}}(\vex{r})$ 
is a ``renormalized and normalized'' LDOS 
moment eigenoperator, 
defined by Eq.~(\ref{LDOSEigenOpsSummedNorm})
in the technical Sec.~\ref{sec: model renomaliation and composite operator renomalization}
of this paper.
[Note that here $pq$ denotes the product of the integers
$p$ and $q$.
$\NoOp{\mathcal{O}_{m}}(\vex{r})$ is just a normalized
version of the LDOS moment operator $\mathcal{O}_{m}(\vex{r})$,
defined previously via Eq.~(\ref{LDOSEigenOpsSummed}). The careful
normalization of operators is an important technical step required
for the accurate computation of correlation functions at the MIT,
as detailed in Sec.~\ref{sec: model renomaliation and composite operator renomalization}.
In this Section, we merely assert that the proper procedure has been implemented.]
Eq.~(\ref{FGenNLsM}) generalizes Eq.~(\ref{LDOSEigenOpsAction}) for 
the case of $q > 1$: in order to compute $F_{q}(\xi; L)$, one must
augment the bare sigma model action with the operator tower 
$\{\NoOp{\mathcal{O}_{q}},\NoOp{\mathcal{O}_{2q}},\NoOp{\mathcal{O}_{3q}},\ldots\}$,
since through the OPE operators representing lower integral LDOS moments generate new ones 
representing higher integral multiples of these. For $q > 1$, the operators 
in Eq.~(\ref{FGenNLsM}) form a \emph{subset} of those in Eq.~(\ref{LDOSEigenOpsAction}). 
The expectation $\langle\cdots\rangle$ in Eq.~(\ref{FGenNLsM}) is taken with respect to the
NL$\sigma$M action at the MIT in $d = 2 + \epsilon$, Eq.~(\ref{SDef}), with $h = 0$ 
and $t = t^{*}$. 
The coupling constants $Y_{p q}$ take the bare values
\begin{equation}\label{BareCouplings}
	Y_{p q}(l = 0) = -\xi \, \delta_{p,1}. 
\end{equation} 
Here, $l = \ln L/L_{0}$ is the log of the spatial length scale $L$ (e.g., the system size),
with $L_{0}$ an arbitrary reference scale.

The simplest approximation to $F_{q}(\xi; L)$ obtains from the lowest
order cumulant expansion of Eq.~(\ref{FGenNLsM}), evaluated at $L = L_{0}$
[i.e.\ using the bare coupling constants in Eq.~(\ref{BareCouplings})]: 
\begin{align}\label{FGenNLsMCumExp}
	F_{q}(\xi; L_{0}) 
	&\sim 
	\exp\left[ 
	-\xi
	\int \dvex{r} \,
	\left\langle
	\NoOp{
	\mathcal{O}_{q}
	}
	\right\rangle
	\right]
	\nonumber\\
	&\sim
	\exp\left(
	-\xi L_{0}^{d - x^{*}_{q}}
	\right).
\end{align}
In this equation, $x^{*}_{q}$ denotes the negative scaling
dimension of the operator $\NoOp{\mathcal{O}_{q}}(\vex{r})$ at the MIT;
for the unitary class studied here, the result to lowest order in $\sqrt{\epsilon}$
was given by 
Eqs.~(\ref{AvgMFCSpecxq}) and (\ref{XiDef}), above.
Combining Eqs.~(\ref{FGenNLsMCumExp}) and (\ref{tau(q)FromFGen}), we immediately recover
Eq.~(\ref{TypISAvg}): the lowest order cumulant approximation to
$F_{q}(\xi; L_{0})$ equates the \emph{typical} $\tau(q)$ spectrum with 
$\tilde{\tau}(q)$ 
[Eqs.~(\ref{AvgMFCSpecI})--(\ref{XiDef})],
associated to the average of the IPR.
For sufficiently large moments 
with $q > q_{c}$, where
$q_{c}$ was defined by Eq.~(\ref{pmqcDef}),
this identification invariably
breaks down (see the discussion in Sec.~\ref{AvgTypTerm}); 
an accurate computation of Eq.~(\ref{FGenNLsMCumExp}) then requires
the retention of higher order cumulants. One immediately sees the need for the
operator product expansion (OPE), as defined by Eq.~(\ref{OPEDef}):
the second and higher cumulants involve products of LDOS moment operators, integrated over the sample
volume. When two (or more) such operators approach the same spatial position,
fusion can occur, in which new, higher moment operators are generated 
through short-distance regularization.\cite{FieldTheory,Cardy96,Carpentier00,Carpentier01,Mudry03}
At the MIT, the negative scaling dimensions $\{x^{*}_{q}\}$ of the LDOS 
moment operators
[Eqs.~(\ref{AvgMFCSpecxq})--(\ref{XiDef})]
satisfy the convexity relation given by Eq.~(\ref{ConvexityIntro}).
Therefore, 
operators corresponding to successively higher moments carry ever more 
negative scaling dimensions, contributing ever more strongly 
to the cumulant expansion. 

Rather than compute the generating 
function
$F_{q}(\xi; L)$
directly, we will use scaling arguments to extract its asymptotic behavior 
in the large system size limit, $L/L_{0} \rightarrow \infty$.
In Eq.~(\ref{FGenNLsM}),
the LDOS moment operators $\NoOp{\mathcal{O}_{p q}}$ perturb the 
action
of the critical field theory. It is well known that the 
lowest order RG equations for the set of conjugate
coupling constants $Y_{p q}$ follow directly from the
operator product expansion.\cite{Cardy96} 

In Sec.~\ref{sec: model renomaliation and composite operator renomalization},
we demonstrate that (the properly normalized versions of) the operators defined 
by Eq.~(\ref{LDOSEigenOpsSummed}) obey the OPE given by Eq.~(\ref{OPEDef}) at 
the MIT in $d = 2 + \epsilon$. We find that the OPE coefficient is given by the ``tree level'' 
(zero coupling) amplitude
\begin{equation}\label{OPECoeffSummary}
	C_{q,q'}^{q+q'}
	=
	\frac{(q + q')!}{q! \, q'!}
	+ \ord{\epsilon}.
\end{equation}
These results are obtained as Eqs.~(\ref{OPECoeff3}) and (\ref{OPEFinal}) in 
Sec.~\ref{sec: model renomaliation and composite operator renomalization},
where we demonstrate that the lowest order $t^{*} \propto \sqrt{\epsilon}$ 
(one-loop) correction to the OPE coefficient in Eq.~(\ref{OPECoeffSummary}) vanishes.
Using Eqs.~(\ref{OPEDef}) and (\ref{OPECoeffSummary}),
one finds the infinite set of RG equations\cite{Cardy96}
\begin{equation}\label{CoupledRGEqs}
	\frac{d Y_{p q}}{d l}
	=
	(d - x^{*}_{p q}) Y_{p q} 
	+ \frac{S_{d}}{2}
	\sum_{m = 1}^{p - 1}
	\binom{p}{m}
	Y_{m q} Y_{(p - m) q}
	+\ord{Y^3}\!,
\end{equation}
where $S_{d}$ is the surface area of the sphere in $d$ dimensions.
Through the OPE, lower moment coupling constants always generate
higher ones; the convexity property in Eq.~(\ref{ConvexityIntro}) implies
that, for $p > p'$, a non-zero $Y_{p q}$ represents 
a much more relevant perturbation than $Y_{p' q}$ 
to the critical NL$\sigma$M action. 
Clearly we must retain the entire infinite set $\{Y_{p q}\}$ in our analysis.

At first glance, the generation of infinitely many relevant couplings
would seem to imply non-universality: there are infinitely many classes of 
solutions to the RG equations (\ref{CoupledRGEqs}), and hence there are 
infinitely many ways to depart from the 
RG fixed point
representing the MIT. This is consistent with the fact that a random critical 
point should be characterized by the entire distribution functions of physical 
quantities, which can become very broad. 
At a delocalization critical point, however,
the multifractal $\tau(q)$ and $f(\alpha)$ spectra, associated to a typical
wavefunction in a fixed disorder realization,
are 
both self-averaging\cite{EversMirlin00,MirlinEvers00}
and universal.\cite{PookJanssen91,Huckestein95,Evers01,Evers2008,Obuse2008} 
We will demonstrate that the functional renormalization group (FRG) method gives
a universal prediction for $\tau(q)$ and $f(\alpha)$, below and above termination (as defined in 
Sec.~\ref{AvgTypTerm}), at the unitary class Anderson MIT in $d = 2 + \epsilon$
consistent with this picture.

We can trade the coupled set of ordinary differential equations in Eq.~(\ref{CoupledRGEqs})
for a single partial differential equation (PDE) by defining the auxiliary generating
function\cite{Mudry03}
\begin{equation}\label{GGenDef}
	G_{q}(z,l) 
	\equiv 
	\tilde{G}_{q}(\tz,\tl)
	\equiv 
	1 + \frac{S_{d}}{2 d} 
	\sum_{p = 1}^{\infty}
	\frac{(e^{-z})^{p}}{p!}
	Y_{pq}(l),	
\end{equation}
where we have introduced the ``position coordinate'' $z$.
$\tilde{G}_{q}(\tz,\tl)$ is a ``Galilean boost'' of
$G_{q}(z,l)$, with 
\begin{align}\label{Boost}
	\tz \equiv z + \Xi q l,
	\quad
	\tl \equiv l.
\end{align}
At the unitary class MIT in $d = 2+\epsilon$, the parameter $\Xi$ has
the value given by Eq.~(\ref{XiDef}),
to one-loop order.

Using the RG equations (\ref{CoupledRGEqs}) for the coupling
constants $\{Y_{p q}\}$, and the explicit form of $x^{*}_{p q}$
from 
Eqs.~(\ref{AvgMFCSpecxq})--(\ref{XiDef}),
one can easily show that $\tilde{G}_{q}(\tz,\tl)$
satisfies the following Kolmogorov-Petrovsky-Piscounov (KPP) equation:
\begin{equation}\label{KPP}
	\frac{1}{d} \partial_{\tl} \tilde{G}_{q}
	=
	D_{q} \partial^2_{\tz} \tilde{G}_{q}
	+ \tilde{G}_{q}(\tilde{G}_{q}-1),
\end{equation}
where we have introduced the effective diffusion constant
\begin{equation}\label{DqDef}
	D_{q} \equiv \frac{q^2 \Xi}{d}.
\end{equation}
The same Eq.~(\ref{KPP}) was obtained in previous FRG studies of
2D disordered systems.\cite{Carpentier00,Carpentier01,Mudry03}

\subsection{Solution to the KPP equation and results}

The KPP equation (\ref{KPP}) describes non-linear diffusion phenomena.
The positive $D_{q}$ in Eq.~(\ref{DqDef}) reflects
the diffusion of the distribution function for the inverse participation
ratio (IPR), defined by Eq.~(\ref{IPRDef}). For high moments,
$q \gg 1$, this diffusion constant is very large, indicating that the IPR
becomes broadly distributed in the large system size limit;
in this regime, the $\tilde{\tau}(q)$ spectrum 
associated with the average
IPR [Eq.~(\ref{tauavg})]
is dominated by 
rare realizations of the disorder,
and loses its meaning with respect to
the typical wavefunction. The non-linear term in Eq.~(\ref{KPP}) appears
because the generating 
functions
$\tilde{G}_{q}(\tz,\tl)$ and $F_{q}(\xi; L)$
encode information about the typical $\tau(q)$ spectrum; this non-linearity
arises 
through
the OPE between LDOS moment operators 
[Eqs.~(\ref{OPEDef}) and (\ref{OPECoeffSummary})].
As explained in the paragraph following Eq.~(\ref{FGenNLsMCumExp}),
the OPE is the essential ingredient required in the computation of $\tau(q)$, 
which was missed in previous treatments\cite{Wegner80,AltshulerKratsovLerner91,AltshulerKratsovLerner86-89}
of the Anderson MIT based upon the NL$\sigma$M approach. 

Non-linear PDEs are often not analytically solvable, but a number of key 
results are known for the KPP equation. We summarize here only those features essential
to the computation of $\tau(q)$; 
for further details,
consult Refs.~\onlinecite{Carpentier00,Carpentier01,Mudry03} and the references
therein.
For a large class of initial conditions which satisfy
\begin{subequations}\label{ICforKPP}
\begin{align}
	\lim_{\tz\rightarrow+\infty}
	\tilde{G}_{q}(\tz,0) &= 1,
	\label{ICforKPP+}\\
	\lim_{\tz\rightarrow-\infty}
	\tilde{G}_{q}(\tz,0) &= 0,
	\label{ICforKPP-}
\end{align}
\end{subequations}
$\tilde{G}_{q}(\tz,\tl)$ converges to a stable traveling wave solution propagating
in the positive $\tz$ direction,
\begin{equation}\label{Travelingwave}
	\tilde{G}_{q}(\tz,\tl\rightarrow\infty)
	\rightarrow
	h(\tz -\tilde{c}_{q} l),
\end{equation}
where the constant $\tilde{c}_{q}$ denotes the wavefront velocity.
The functional form of the traveling wave in Eq.~(\ref{Travelingwave}) 
is sensitive to the details of the initial condition at $\tl = 0$. 
On the contrary,
for an initial wavefront satisfying the asymptotic property
\begin{equation}\label{KPPICAsym}
	\tilde{G}_{q}(\tz\rightarrow+\infty, 0)
	\sim
	1 - \lambda e^{-\tz},
\end{equation}
with $\lambda$ a pure number, the velocity $\tilde{c}_{q}$ is \emph{universal}, 
depending only upon the diffusion constant $D_{q}$, defined in the context 
of the MIT by Eq.~(\ref{DqDef}), above.
Note that Eq.~(\ref{KPPICAsym}) constrains $\tilde{G}_{q}$ only in the region
penetrated by the wavefront [Eq.~(\ref{Travelingwave})] in the limit of 
large ``renormalization time,'' $\tl \rightarrow \infty$.
Remarkably, the wavefront velocity is also insensitive to the precise 
form of the non-linear term $\mathcal{F}(\tilde{G}_{q}) \equiv \tilde{G}_{q} (\tilde{G}_{q} - 1)$ 
in the KPP equation (\ref{KPP}). In fact, the same velocity obtains from KPP for
any nonlinear forcing function satisfying the constraints
\begin{align}\label{ForcingFunctions}
	\mathcal{F}(0) = \mathcal{F}(1) = 0,\qquad
	\mathcal{F}(\tilde{G}) < 0,
	\nonumber\\
	\frac{d\mathcal{F}(\tilde{G})}{d \tilde{G}} \geq -1,\qquad
	\frac{d\mathcal{F}(\tilde{0})}{d \tilde{G}} = -1,
\end{align}
for $0 \leq \tilde{G} \leq 1$. 
In this sense, the KPP equation achieves a strong version of universality.

Let us now return to the problem at hand, computing the typical $\tau(q)$
spectrum obtained at the MIT in the unitary class 
for $d = 2 + \epsilon$.
The initial condition for the KPP Eq.~(\ref{KPP}) implied
by Eq.~(\ref{BareCouplings}) is
\begin{equation}\label{BareICforKPP}
	\tilde{G}_{q}(\tz, 0)
	=
	1 -\xi \frac{S_{d}}{2 d} e^{-\tz},
\end{equation}
consistent with only $Y_{q}$ non-vanishing. Eq.~(\ref{BareICforKPP}) 
satisfies the condition in Eq.~(\ref{ICforKPP+}), having the same form as that expressed
in Eq.~(\ref{KPPICAsym}). In order to satisfy Eq.~(\ref{ICforKPP-}), we must bound
the amplitude $0 \leq \tilde{G}_{q}(\tz\rightarrow -\infty,0) \leq 1$; 
to that end, we deform Eq.~(\ref{BareCouplings}) as follows:
\begin{align}\label{BareCouplingsDeformed}
	Y_{q}(0) &= -\xi,
	\nonumber\\
	Y_{p q}(0) 
	&\rightarrow
	\left(\frac{S_{d}}{2 d}\right)^{p-1} \,\left[Y_{q}(0)\right]^p,
\end{align}
which leads to
\begin{equation}\label{BareICforKPPDeformed}
	\tilde{G}_{q}(\tz,0) 
	\sim
	\exp\left[
	- \xi \frac{S_{d}}{2 d} e^{-\tz}
	\right].
\end{equation}

Crucially, since Eq.~(\ref{BareICforKPPDeformed}) satisfies Eq.~(\ref{KPPICAsym}), 
the asymptotic traveling wave velocity $\tilde{c}_{q}$ [Eq.~(\ref{Travelingwave})] 
depends only upon the diffusion constant $D_{q}$, Eq.~(\ref{DqDef}).
For the KPP equation (\ref{KPP}) satisfying 
(\ref{ICforKPP+}), (\ref{ICforKPP-}), and (\ref{KPPICAsym}),
one finds qualitatively different behavior for $D_{q}$ less than or greater than
one:\cite{Carpentier00,Carpentier01,Mudry03}
\begin{equation}\label{KPPTildeVelocity}
	\tilde{c}_{q} =
	\left\{
	\begin{aligned}
	&d\left(1 + D_{q}\right), && D_{q} \leq 1
	\\	
	&2 d \sqrt{D_{q}}, && D_{q} > 1
	\end{aligned}
	\right.
\end{equation}
Reversing the Galilean boost in Eq.~(\ref{Travelingwave}) via
Eq.~(\ref{Boost}), we see that
\begin{align}\label{KPPVelocity}
	G_{q}(z,l \rightarrow \infty)
	&\sim 
	h\left(
	z - c_{q} l
	\right), 
	\nonumber\\
	c_{q} 
	&\equiv
	\tilde{c}_{q} - \Xi q.
\end{align}

Let us try to understand the physics implied by Eq.~(\ref{KPPVelocity}).
The generating 
function
$G_{q}(z,l)$ was defined via Eq.~(\ref{GGenDef}) 
in terms of the infinite tower of coupling constants $\{Y_{pq}\}$; the latter
were introduced in the NL$\sigma$M definition of $F_{q}(\xi;L)$, Eq.~(\ref{FGenNLsM}). 
Under a change of length scale (e.g.\ an incremental increase in the sample size), each $Y_{pq}$ 
evolves according to the RG Eq.~(\ref{CoupledRGEqs}). If we neglect the non-linear
terms in this equation due to the OPE, then each non-zero $Y_{pq}$ grows under
renormalization according to its own scaling exponent $c_{pq}^{(0)} \equiv d - x^{*}_{pq}$; 
a full characterization of the system requires the specification of the entire set 
$\{c_{pq}^{(0)}\}$, $p \in \mathbb{N}$. As argued in Sec.~\ref{Intro} and below 
Eq.~(\ref{CoupledRGEqs}), we nevertheless expect that a single, well-defined 
exponent $\tau(q)$ can be defined for the IPR associated to a typical wavefunction, even in the limit
of 
relatively ``large'' $q$. 
Eq.~(\ref{KPPVelocity}) implies that, through the OPE and
the subsequent non-linearity of the KPP equation, the functional RG proves
this assertion: the asymptotic scaling of $G_{q}(z,l)$ involves a single number,
the velocity $c_{q}$ given by Eqs.~(\ref{KPPTildeVelocity}) and (\ref{KPPVelocity}),
which we can think of as a ``typical'' scaling exponent. 

Since $G_{q}(z,l)$ tracks the scaling of coupling constants, we infer that the 
set $\{Y_{pq}\}$ ``fuses'' into a single, typical coupling $Y_{q}^{\mathsf{typ}}$, up
to less relevant perturbations to the critical NL$\sigma$M fixed point; 
we can then define an associated typical anomalous dimension
\begin{align}\label{ScalingDimTyp}
	x_{q}^{\mathsf{typ}} 
	&\equiv 
	d - c_{q}
	\nonumber\\
	&=
	\left\{
	\begin{aligned}
	&-\Xi q(q-1), 
	&& 1 \leq q \leq q_{c} 
	\\	
	&
	d(1-q) 
	+ q \left(\sqrt{d} - \sgn(q) \sqrt{\Xi}\right)^2, 
	&& q > q_{c}
	\end{aligned}
	\right.
\end{align}
where we have used Eq.~(\ref{DqDef}).
In this equation, the critical value 
$q_{c}$
was 
defined in the Introduction, Eq.~(\ref{pmqcDef}). 

We infer from Eq.~(\ref{ScalingDimTyp}) that 
$F_{q}(\xi;L)$ 
acquires the following asymptotic form:
\begin{align}\label{FGenNLsMTyp1}
	F_{q}(\xi; L\rightarrow \infty) 
	&\sim 
	\left\langle 
	\exp\left[ 
	Y_{q}^{\mathsf{typ}}
	\int \dvex{r} \,
	\NoOp{
	\mathcal{O}_{q}^{\mathsf{typ}}
	}
	(\vex{r})
	\right]
	\right\rangle
	\nonumber\\
	&\sim
	\exp\left(
	Y_{q}^{\mathsf{typ}} L^{d - x^{\mathsf{typ}}_{q}}
	\right).
\end{align}
As in Eq.~(\ref{FGenNLsMCumExp}), we have evaluated $F_{q}(\xi; L)$ 
in the lowest order cumulant expansion; the crucial difference
between Eqs.~(\ref{FGenNLsMCumExp}) and Eq.~(\ref{FGenNLsMTyp1}) resides
in the implied order of operations. To obtain the final result in Eq.~(\ref{FGenNLsMTyp1}),
we first coarse grain the system, say by integrating-out short wavelength
degrees of freedom (in a Wilsonian picture).
The coarse graining generates higher order couplings $\{Y_{pq}\}$, $p>1$, 
through the non-linear RG Eq.~(\ref{CoupledRGEqs}). In the large system
size limit $l = \ln(L/L_{0}) \rightarrow \infty$, a single, well-defined
typical coupling $Y_{q}^{\mathsf{typ}}$ emerges, associated to a new
local operator $\NoOp{\mathcal{O}_{q}^{\mathsf{typ}}}(\vex{r})$,
whose scaling dimension is given by Eq.~(\ref{ScalingDimTyp}).
Finally, we evaluate $F_{q}(\xi; L\rightarrow \infty)$ to lowest order in the
cumulant expansion, which gives Eq.~(\ref{FGenNLsMTyp1}). 
This is expected to be 
a correct representation of the
the asymptotic scaling limit,
because the functional RG has already built all of the most relevant
operator ``fusions'' into the definition of $\NoOp{\mathcal{O}_{q}^{\mathsf{typ}}}(\vex{r})$.
The emergence of the associated $Y_{q}^{\mathsf{typ}}$ and $x^{\mathsf{typ}}_{q}$ has
been proven above using the properties of the KPP equation, Eq.~(\ref{KPP}).

Finally, we extract the typical $\tau(q)$ spectrum.
As obtained in the limit of large, but finite renormalization, the typical 
coupling $Y_{q}^{\mathsf{typ}}$ should have an analytic expansion
in powers of the parameter $\xi$ [c.f.\ Eqs.~(\ref{BareCouplings}) and 
(\ref{BareCouplingsDeformed})]. Up to an irrelevant rescaling, we may write
\begin{equation}\label{YqtypExp}
	Y_{q}^{\mathsf{typ}} = -\xi - \sum_{m = 2}^{\infty} \mathcal{Y}_{q m}^{\mathsf{typ}} \xi^{m}.
\end{equation}
Combining Eqs.~(\ref{tau(q)FromFGen}) and (\ref{FGenNLsMTyp1}), we obtain
\begin{equation}\label{tau(q)typFinalA}
	\tau(q) 
	\sim
	\frac{d\phantom{\ln}}{d \ln L}
	\int_{0}^{\infty} \frac{d \xi}{\xi} 
	\left[
	e^{Y_{q}^{\mathsf{typ}} L^{d - x^{\mathsf{typ}}_{q}}} 
	- q \, e^{Y_{1}^{\mathsf{typ}} L^{d - x^{\mathsf{typ}}_{1}}} 
	\right].
\end{equation}
In the limit $L \rightarrow \infty$, we may neglect all but
the first term in Eq.~(\ref{YqtypExp}), since $d - x_{q}^{\mathsf{typ}} \geq 0$ 
for all $q \geq 1$, provided $\Xi \leq  4 d$. 
This condition is always satisfied in the perturbatively accessible regime, $0<\epsilon \ll 1$, 
where the parameter 
$\Xi = \sqrt{\epsilon/2} + \ord{\epsilon} \ll 1$ [Eq.~(\ref{XiDef})].
Then we obtain, using Eq.~(\ref{LogIntIdent})
\begin{align}\label{tau(q)typFinalB}
	\tau(q) =& 
	d(q - 1) 
	+ x_{q}^{\mathsf{typ}} - q \, x_{1}^{\mathsf{typ}}.
\end{align}
Eq.~(\ref{tau(q)typFinalB}) for the typical $\tau(q)$ should be compared to 
Eq.~(\ref{AvgMFCSpecI}) for $\tilde{\tau}(q)$.
In the perturbative regime $\epsilon \ll 1$, 
we have $x_{1}^{\mathsf{typ}} = 0$ [Eq.~(\ref{ScalingDimTyp})]. Combining
Eqs.~(\ref{ScalingDimTyp}) and (\ref{tau(q)typFinalB}), we arrive
at our final result, the typical $\tau(q)$ spectrum given by
\begin{align}\label{tau(q)typFinalC}
	\tau(q) =& 
	\left\{
	\begin{aligned}
	&d(q - 1)\left(1 - \frac{q}{q_{c}^2}\right), && |q| \leq q_{c}
	\\	
	& d \left(1 - \frac{\sgn(q)}{q_{c}}\right)^2 q, && |q| > q_{c}
	\end{aligned}
	\right.
\end{align}
where $q_{c} = \sqrt{{d}/{\Xi}}$ [Eq.~(\ref{pmqcDef})].
In this equation, 
we have extended $q$ from the positive integers to
the entire real line.
The average $\tilde{\tau}(q)$ and typical $\tau(q)$ spectra are respectively sketched 
in the top and bottom panels of Fig.~\ref{FigTauq} in Sec.~\ref{Intro}.

The singularity spectrum $f(\alpha)$ was introduced in Eq.~(\ref{fDef}).
The $f(\alpha)$ corresponding to the typical $\tau(q)$ in Eq.~(\ref{tau(q)typFinalC})
is
\begin{align}\label{TypfSpec}
	f(\alpha)
	&=
	\left\{
	\begin{aligned}
	&d - \frac{(\alpha - d - \Xi)^2}{4 \Xi} &&
	\\
	&\quad=\frac{q_{c}^2(\alpha_{+} - \alpha)(\alpha -\alpha_{-})}{4 d}, && \alpha_{-} \leq \alpha \leq \alpha_{+}
	\\	
	& 0, && \alpha<\alpha_{-},\,\alpha_{+}<\alpha
	\end{aligned}
	\right.
\end{align}
The spectral cutoffs $\alpha_{\pm}$ were defined by Eq.~(\ref{alphapmDef}).
As expected, $f(\alpha)$ associated to the typical wavefunction is 
never negative, as discussed in Sec.~\ref{Intro};
see also the top and bottom panel insets in Fig.~\ref{FigTauq}.
Eqs.~(\ref{tau(q)typFinalC}) and (\ref{TypfSpec}) hold to the lowest non-trivial
order in $\sqrt{\epsilon}$.
The consistency of the restriction to only the lowest order contributions
in the $\epsilon$ expansion is
demonstrated in Sec.~\ref{sec: conclusion}.

An alternative representation of multifractality invokes the ``generalized dimension''
$D_{q}$, defined via
\begin{equation}\label{GenDim}
	\tau(q) \equiv (q - 1)D_{q}.
\end{equation}
Spectral termination [Eqs.~(\ref{tau(q)typFinalC}) and (\ref{TypfSpec})] implies that 
$\alpha_{+} \leq D_{q} \leq \alpha_{-}$, with the boundary values associated to
the limits 
\begin{equation}\label{GenDimBounds}
	\lim_{q \rightarrow \pm \infty} D_{q}
	= \alpha_{\mp}.
\end{equation}
Numerical computations of $D_{q}$ for the typical wavefunction confirm Eq.~(\ref{GenDimBounds}); 
see, e.g., 
Refs.~\onlinecite{SchreiberGrussbach91} and \onlinecite{PookJanssen91}.

\subsection{Comparison with other arguments}\label{ComparisonOtheMethods}

Our final results (\ref{tau(q)typFinalC}) and (\ref{TypfSpec}) agree with 
previous heuristic arguments given in Refs.\ \onlinecite{Castillo97}, \onlinecite{EversMirlin00},
and \onlinecite{MirlinEvers00}. As explained below Eq.~(\ref{fDef}), $f(\alpha)$ describes 
the measure $L^{f(\alpha)}$ of the set of those points $\vex{r}$ where the eigenfunction $\psi$ 
takes the value $|\psi(\vex{r})|^2 \propto L^{-\alpha}$.\cite{Halsey86}
Hence, the IPR $P_q$ [Eq.~(\ref{IPRDef})] can be estimated as an integral
\begin{eqnarray}
	P_q \sim
	\int_{f(\alpha)\ge 0}
	d\alpha\,
	L^{-q\alpha+f(\alpha)}.
\end{eqnarray}
The integrand takes a maximum value at 
a saddle point value $\alpha$, which defines 
the $\tau(q)$.
For $|q|<q_c$, this saddle point is in the integration
domain, whereas for $|q|>q_c$ it is outside of
it.
In the latter case, the integral is dominated by the boundary
value of $\alpha = \alpha_{\mp}$, where $f(\alpha_{\mp})=0$.


\section{Discussion} 
\label{sec: conclusion}

We have provided a field theoretical description
of the termination
of the multifractal spectrum $\tau(q)$, 
as defined for the typical wavefunctions, at the Anderson MIT in $d = 2 + \epsilon$ for 
the unitary disordered metal class. The essential ingredients of the calculation are 
evident in the formulation of Eqs.~(\ref{tau(q)FromFGen}) and (\ref{FGenNLsM}): 
these are the infinite set of properly normalized LDOS moment operators $\{\NoOp{\mathcal{O}_{p q}}(\vex{r})\}$,
$p \in \mathbb{N}$,
characterized by the negative scaling dimensions $\{x^*_{p q}\}$ in 
Eqs.~(\ref{AvgMFCSpecxq})--(\ref{XiDef}).
Each successive higher moment operator with $p = \{1,2,\ldots\}$ constitutes a more 
strongly relevant perturbation to the critical RG fixed point 
that describes the MIT.
Through the OPE [Eq.~(\ref{OPEDef})], lower moments
always generate higher ones,
and a consistent treatment of the problem requires that the entire 
infinite hierarchy of LDOS moment operators is retained.
Ordinarily, the advent of an infinity of relevant scaling directions should cast 
serious doubt upon the adequacy of single (or few) parameter scaling, at least with
respect to the investigated critical point;
remarkably, the FRG 
``absorbs''
the entire LDOS moment tower, and through the 
(universal) properties of the long-time asymptotics 
of the KPP equation, 
renders in the end a single, \emph{universal} prediction for the typical $\tau(q)$. 

Physically, the relevant LDOS moments reflect the fact that a random critical point 
should be characterized by the distribution functions of physical quantities, rather than 
their mean, variance, or first few moments. The distribution of an observable
in the presence of quenched disorder can become very broad, due to the influence of rare 
events.\cite{AltshulerKratsovLerner86-89,AltshulerKratsovLerner91,Muzykantskii94,Falko95,Mirlin95}
In principle, we need the functional renormalization group (FRG) to obtain scaling for
the entire probability distribution.\cite{Yudson}
For large $q$, the IPR, defined by Eq.~(\ref{IPRDef}) [or of its field theoretic generalization, 
Eq.~(\ref{IPRtoLDOS})], constitutes such a broadly-distributed 
observable.\cite{Fyodorov95,Chamon96,PrigodinAltshuler98,MirlinReview1and2,EversMirlin00,MirlinEvers00} 
By comparison, a universal $\tau(q)$ spectrum for the typical wavefunction obtains because 
the log of the IPR is self-averaging for all $q$.\cite{EversMirlin00,MirlinEvers00}

Technically, the
FRG method implemented in Sec.~\ref{sec: FRG} is completely analogous to that
employed previously\cite{Carpentier00,Carpentier01,Mudry03} in the study of certain 
special 2D disordered field theories, possessing an additional, `chiral' symmetry.\cite{DisClasses}
[See the end of 
Sec.~\ref{IntroOPE} 
for a description of these chiral models.]
As in this prior work, the FRG translates 
the infinite set of coupled
flow equations (\ref{CoupledRGEqs}) into the KPP Eq.~(\ref{KPP}) for a
certain (auxiliary) generating 
function.
Through the asymptotic solution of the KPP equation in the form of 
a propagating 
wavefront,
the tower of relevant LDOS moment operators
combines via multiple  OPEs
into a single, 
typical operator (up to less relevant
perturbations), characterized by the typical scaling dimension in Eq.~(\ref{ScalingDimTyp}).
The final results for $\tau(q)$ and $f(\alpha)$ [Eqs.~(\ref{tau(q)typFinalC}) 
and (\ref{TypfSpec})] are obtained via the FRG for the unitary universality class,
using only two inputs, evaluated at the MIT: {\bf(i)} the scaling dimensions $\{x^*_q\}$
(associated to the \emph{average} operator scaling, already known from previous 
work)\cite{Wegner80,Pruisken85,HofWegner86,Wegner87} and {\bf(ii)} the OPE coefficient 
$C_{q,q'}^{q+q'}$, 
Eq.~(\ref{OPECoeffSummary}) 
(computed in Sec.~\ref{sec: model renomaliation and composite operator renomalization}
to lowest non-trivial order in $\sqrt{\epsilon}$).
The former is specific to the unitary class, but 
we have shown that the latter takes exactly
the same form in the chiral model calculations.\cite{Carpentier00,Carpentier01,Mudry03}

In treating the unitary class, we have chosen to work only to 
the 
lowest order in $t^* \propto \sqrt{\epsilon}$,
i.e.\ to one loop. 
To this order, the
resulting singularity spectrum $f(\alpha)$ given by Eq.~(\ref{TypfSpec}) 
is purely quadratic over the region $\alpha_{+} < \alpha < \alpha_{-}$ (the so-called ``parabolic 
approximation'').\cite{Janssen94,Huckestein95,Mildenberger01,Evers2008,Obuse2008}
We now discuss the consistency of working with the functional renormalization
group to this order in the
$\epsilon$-expansion.
Corrections to the LDOS moment scaling dimensions $\{x^*_q\}$ are already known to four 
loops,\cite{HofWegner86,Wegner87}
\begin{align}\label{LDOSScalingDimHOT}
	x^*_q 
	&=
	- \sqrt{\frac{\epsilon}{2}}q(q-1)
	- \frac{3\,\zeta(3)}{8} \epsilon^2 q^2 (q-1)^2
	+\ord{\epsilon^{5/2}},
\end{align}
where $\zeta(z)$ denotes the Riemann zeta function.
While the FRG method formally retains LDOS moment operators $\{\NoOp{\mathcal{O}_{q}}(\vex{r})\}$
to arbitrarily high orders in $q$, it is crucial to note that the termination
of the typical $\tau(q)$ spectrum [Eq.~(\ref{tau(q)typFinalC})] occurs at the finite value $q = q_{c}$
[Eq.~(\ref{pmqcDef})]; to lowest order,
\begin{equation}\label{qcExpansion}
	q_c^2 
	=
	2 \sqrt{2/\epsilon} + \ord{1}.
\end{equation}
Evaluating the 4-loop scaling dimension in Eq.~(\ref{LDOSScalingDimHOT}) at $q_{c}$, we obtain
\begin{align}\label{LDOSScalingDimHOTqc}
	x^*_{q = q_{c}} 
	=
	- 2
	+ \left(2 \epsilon\right)^{1/4}
	+ \ord{\epsilon^{1/2}}.
\end{align}
The one-loop approximation consists of retaining 
only the first term on the right-hand side (RHS) of Eq.~(\ref{LDOSScalingDimHOT}), as well as
the terms written explicitly on the RHS of each of Eqs.~(\ref{qcExpansion}) and (\ref{LDOSScalingDimHOTqc}).
At termination ($q = q_{c}$), the higher order loop corrections give rise to 
additional terms in Eq.~(\ref{LDOSScalingDimHOTqc}) that are down by 
higher powers of $\epsilon^{1/4}$, and these can be consistently neglected for $\epsilon \ll 1$.

We cannot resist contemplating, at a very speculative level,
a naive extrapolation of our one-loop results to moderate or even large $\epsilon$.
First, note that Eq.~(\ref{ScalingDimTyp}) implies the existence of a
non-zero, typical scaling dimension $x^{\mathsf{typ}}_{1}$ for the first moment of
the LDOS, when $q_{c} < 1$: 
\begin{equation}\label{ScalingDimTypDOS}
	x_{1}^{\mathsf{typ}} 
	=
	d \left(1 - \frac{1}{q_{c}}\right)^2, 
	\quad q_{c} < 1.
\end{equation}
Using the lowest order result in Eq.~(\ref{qcExpansion}),
we then define 
\begin{align}\label{FreezingEpsilon}
	\epsilon_{\mathsf{F}}
	&\equiv 
	\epsilon(q_{c} = 1)
	\nonumber\\
	&\sim 8.
\end{align}
At face value, a $x^{\mathsf{typ}}_{1} > 0$ would imply that
the typical
LDOS
\emph{vanishes} at the MIT. 
Eq.~(\ref{FreezingEpsilon}) suggests that this 
becomes possible in the limit of large spatial dimensionality,
$\epsilon > \epsilon_{\mathsf{F}}$. 
Such a scenario does not contradict rigorous results\cite{Wegner81,McKaneStone81} 
which prove that the average, global
DOS remains uncritical (constant) across the transition
for any spatial dimension $d$.
Indeed, Bethe lattice computations\cite{Efetov97} exhibit a 
typical LDOS that
vanishes \emph{exponentially} across the transition.\cite{TypDOSVanish}
As the Bethe lattice can be equated with the limit of infinite spatial dimensionality,\cite{MirlinReview3}
this picture in fact appears consistent with Eqs.~(\ref{ScalingDimTypDOS})
and (\ref{FreezingEpsilon})
in the limit $\epsilon \rightarrow \infty$, where upon $x^{\mathsf{typ}}_{1} \rightarrow \infty$.

It has been asserted\cite{MirlinReview3} that the upper critical dimension for Anderson localization
occurs only at $d = \infty$, i.e., the Bethe lattice case. The naive extrapolation
of the results of this paper to large $\epsilon$ suggests an alternate possibility. 
To motivate the basic underlying idea, we note
that for $\epsilon > \epsilon_{\mathsf{F}}$, the typical multifractal spectrum
$\tau(q)$ defined by Eqs.~(\ref{ScalingDimTyp}) and (\ref{tau(q)typFinalB})
would take the form
\begin{align}\label{tau(q)typFrozen}
	\tau(q) =& 
	\left\{
	\begin{aligned}
	&-d \left(1 - \frac{q}{q_{c}}\right)^2, && |q| \leq q_{c}
	\\	
	& \frac{2 d}{q_{c}} \left(q - |q|\right). && |q| > q_{c}
	\end{aligned}
	\right.
\end{align}
This should be contrasted with Eq.~(\ref{tau(q)typFinalC}), which assumed
$q_{c} > 1$ [always the case in the perturbatively accessible regime, 
$0 < \epsilon \ll 1$---see Eq.~(\ref{qcExpansion})]. Eq.~(\ref{tau(q)typFrozen}) 
shows that $\tau(q) = 0$ for all $q \geq q_{c}$ when $q_{c} < 1$ ($\epsilon > \epsilon_{\mathsf{F}}$).
Thus the $\tau(q)$ spectrum ``freezes'' for dimensionalities above
the threshold $d_{\mathsf{F}} \equiv 2 + \epsilon_{\mathsf{F}}$. An analogous freezing transition 
has been predicted\cite{Chamon96,Castillo97,Carpentier00,Carpentier01,Motrunich02,Mudry03} 
for the 2D chiral Dirac models discussed at the end of Sec.~\ref{AvgTypTerm};
in these models, the transition occurs for quenched disorder fluctuation strengths 
larger than some threshold value. Unlike the unitary metal class discussed here, the chiral
model freezing transition has been rigorously derived through strong randomness\cite{Motrunich02}
and 
FRG arguments,\cite{Mudry96,Carpentier00,Carpentier01,Mudry03}
which do not rely upon expansion in a small parameter. By comparison, 
Eqs.~(\ref{ScalingDimTypDOS}) and (\ref{tau(q)typFrozen}) lie well beyond the
perturbatively accessible regime. Regardless, the freezing scenario suggests
the intriguing possibility of a finite $d_{\mathsf{F}} < \infty$ for Anderson localization
in the normal metal classes, which could perhaps serve as a finite upper critical 
dimension.


\section{
Operator 
product expansion
at the Anderson fixed point: perturbative calculation
\label{sec: model renomaliation and composite operator renomalization}}

In this final (technical) section, we provide a derivation of the LDOS moment operator algebra
required for the functional RG construction in Sec.~\ref{sec: FRG}.
Using the NL$\sigma$M framework reviewed in Sec.~\ref{sec: definitions and model},
we first rederive the anomalous scaling dimensions of the LDOS moment operators
introduced in Sec.~\ref{sec: tildetau spectrum}. We then turn to the perturbative 
evaluation of the operator product expansion, as defined by Eq.~(\ref{OPEDef}), 
between properly normalized versions of these eigenoperators.

\subsection{Renormalization of the model}\label{TECH1--ModelRenorm}

\begin{figure}
\includegraphics[width=0.2\textwidth]{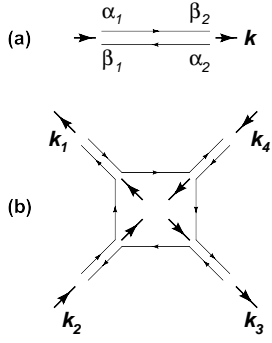}
\caption{Propagator (a) and lowest-order vertex (b) necessary for the
one-loop RG, obtained by expanding Eq.~(\ref{SDef}) in terms of the unconstrained $\Wh$
field, using Eq.~(\ref{WDef}). Associated amplitudes are given by Eqs.~(\ref{Prop}) and (\ref{Vertex})
in the text.
\label{FigRules}}
\end{figure}

To begin, we consider the renormalization of the bare NL$\sigma$M defined by Eq.~(\ref{SDef}).
This calculation is standard;\cite{Wegner79,BrezinHikamiZinnJustin80,LeeRamakrishnan} we provide only our
conventions necessary to set up the computation, and the corresponding results. Using the parameterization
in Eq.~(\ref{WDef}), one obtains the $\Wh\rightarrow W^{\alpha}{}_{\beta}$ field propagator and vertex 
shown in Fig.~\ref{FigRules}. 
The propagator is pictured in Fig.~\ref{FigRules}(a) as pair of 
counter-directed 
thin lines, representing physically 
an ambulating electron-hole pair 
(`diffuson'), and 
mathematically the linking of direct and conjugate indices in two inequivalent representations of U($n$)
[since the maximum compact subgroup of
$\mathrm{U}(2n)$
is U($n$)$\times$U($n$)]. Equivalently, in the unitary class 
with this parameterization, the field $W^{\alpha}{}_{\beta}$ is
Wick-contracted only with its adjoint 
$W^{\dagger\, \beta}{}_{\alpha}$; this fact is indicated by the thick arrows in Fig.~\ref{FigRules}, 
which also encode the direction of momentum flow. The amplitude corresponding to the propagator in 
Fig.~\ref{FigRules}(a) is 
\begin{equation}\label{Prop}
	\langle W^{\alpha_1}{}_{\beta_1}(\vex{k}) W^{\dagger\,\alpha_2}{}_{\beta_2}(\vex{k})\rangle
	=
	\delta^{\alpha_1}_{\beta_2} \delta^{\alpha_2}_{\beta_1} 
	\frac{t_{0}}{|\vex{k}|^{2} + h_{0} t_{0}},
\end{equation}
where $t_{0}$ and $h_{0}$ are bare parameters.
In this paper, we will only need the lowest order non-linear vertex $\equiv\mathfrak{V}_{4}$ 
[obtained via an expansion 
of
Eq.~(\ref{SDef}) in powers of $\Wh$]; this vertex is pictured in 
Fig.~\ref{FigRules}(b), with the corresponding amplitude 
\begin{equation}\label{Vertex}
	\mathfrak{V}_{4}
	=
	\frac{2!}{8 t_{0}}\left[
	\begin{aligned}
	&2(\vex{k_1}\cdot\vex{k_3}+\vex{k_2}\cdot\vex{k_4}) 
	-\vex{k_1}\cdot\vex{k_2}\\
	&-\vex{k_2}\cdot\vex{k_3}-\vex{k_3}\cdot\vex{k_4}-\vex{k_4}\cdot\vex{k_1} 
	- 2 h_{0} t_{0}
	\end{aligned}
	\right].
\end{equation}

Adopting standard dimensional regularization conventions,\cite{BrezinZinnJustinLeGuillou76,BrezinHikamiZinnJustin80,FieldTheory} 
\begin{align}\label{DimReg}
	\begin{aligned}
	t_{0} &\equiv t \mu^{-\epsilon} F_{t},\\
	h_{0} &\equiv Z_{W}^{-\frac{1}{2}} h,
	\end{aligned}
\end{align}
with $t$ and $h$ 
renormalized parameters, $\mu$ an arbitrary inverse-length scale, and $Z_{W}$ the 
field renormalization of the elementary operator $\Wh$, the RG flow equations are given by
\begin{align}\label{DimRegFlow}
	\begin{aligned}
	\frac{d t}{d l} &= \frac{-\epsilon t}{1+\frac{d \ln F_{t}}{d \ln t}},\\
	\frac{d \ln h}{d l} &=d + \frac{1}{2} \frac{d\ln Z_{W}}{d \ln t}\frac{d \ln t}{dl}.
	\end{aligned}
\end{align}
In Eqs.~(\ref{DimReg}) and (\ref{DimRegFlow}), $d = 2 + \epsilon$ is the spatial dimensionality of the system, 
and $l \sim -\ln\mu$ is the logarithm of the spatial length scale.

For the compact 
unitary model with target space 
$\mathrm{U}(2n)/\mathrm{U}(n)\times\mathrm{U}(n)$,
the one-loop flow equations
are 
\begin{align}
	\frac{d t}{d l} &= -\epsilon t + \frac{n t^{2}}{4\pi} + \ord{t^{3}},\label{tFlow1loop}\\ 
	\frac{d \ln h}{d l} &=d - \frac{n t}{4\pi}+ \ord{t^{2}}.\label{hFlow1loop}
\end{align}
These equations possess a critical fixed point at the 
``temperature'' $t^{*} = 4\pi\epsilon/n$,
with $h = 0$. The Anderson model corresponds to the limit $n \rightarrow 0$ in Eqs.~(\ref{tFlow1loop}) and
(\ref{hFlow1loop}); 
in this case, 
a non-trivial critical point occurs at two loop order.\cite{LeeRamakrishnan} 
In our conventions, the two-loop result is\cite{Wegner79,BrezinHikamiZinnJustin80}
\begin{equation}\label{tFlow2loop}
	\frac{d t}{d l} = -\epsilon t + \frac{t^{3}}{2^{5} \pi^{2}} + \ord{t^{4}},
\end{equation}
valid in the limit $n \rightarrow 0$. The critical value of the inverse conductance at the metal-insulator
transition in $d = 2 + \epsilon$ is proportional to $t^{*} = 4\pi\sqrt{2\epsilon} + \ord{\epsilon}$.

\subsection{Composite operator scaling dimensions}\label{TECH2--ScalingDim}

\begin{figure}
\includegraphics[width=0.4\textwidth]{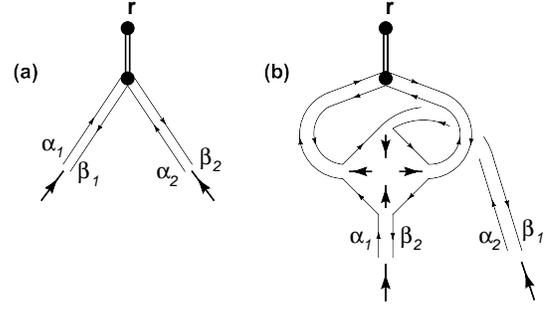}
\caption{Basic renormalization process of a composite operator.
\label{FigScalingDim}}
\end{figure}

Next, we turn to the renormalization of the composite operators defined by Eq.~(\ref{LDOSEigenOps}).
The renormalization of 
$\mathcal{O}^{\; \alpha_{1} \alpha_{2} \ldots \alpha_{p}}_{p \; [\beta_{1} \beta_{2} \ldots \beta_{p}]}$
can be determined by considering ``matrix elements'' of that operator with arbitrary configurations of $p$ 
distant, mutually separated 
adjoint fields $\{W^{\dagger\,\lambda}{}_{\gamma}\}$.
[For more general operators built out of products of both `$\pi$' ($\Wh$,$\Wh^\dagger$) and `$\sigma$' 
($\sqrt{\hat{\mathbb{I}}_{n} - \Wh \Wh^{\dagger}}$,$\sqrt{\hat{\mathbb{I}}_{n} - \Wh^{\dagger} \Wh}$) 
components, one must typically consider multiple matrix element
types involving different numbers of $\Wh$ and $\Wh^{\dagger}$ fields.\cite{BrezinZinnJustinLeGuillou76}]

Specifically, we define
 \begin{multline}\label{MatrixElemDef}
	\Gamma^{(0)\phantom{p} \lambda_{1}\ldots\lambda_{p}}_{p \phantom{(0)}\gamma_{1}\ldots\gamma_{p}}
	[\mathcal{O}^{\; \alpha_{1} \ldots \alpha_{p}}_{p \; [\beta_{1} \ldots \beta_{p}]}(\vex{r})]
	\\
	\equiv \langle\langle 
	\mathcal{O}^{\; \alpha_{1} \ldots \alpha_{p}}_{p \; [\beta_{1} \ldots \beta_{p}]}(\vex{r})\,\,
	W^{\dagger\, \lambda_{1}}{}_{\gamma_{1}} \cdots W^{\dagger\, \lambda_{p}}{}_{\gamma_{p}}
	\rangle\rangle,
\end{multline}	
where the double angle brackets $\langle\langle \mathcal{O} \cdots \rangle\rangle$ signify the 
one-particle irreducible matrix element of $\mathcal{O}$, amputating the external 
fields.\cite{FieldTheory,footnote-VertexDef} 
The external $\{\Wh^{\dagger}\}$ fields are assumed to be 
located far from each other, and from the position $\vex{r}$ of the composite operator. The superscript 
$(0)$ on the left-hand side (LHS) of this equation indicates that this is a bare (i.e.\ not yet 
renormalized) quantity.

The basic one-loop process is illustrated in Fig.~\ref{FigScalingDim}. 
Fig.~\ref{FigScalingDim}(a) depicts the two-field matrix element of the (unsymmetrized) operator
\begin{equation}\label{TwoLegOp}
	W^{\alpha_1}{}_{\beta_1} W^{\alpha_2}{}_{\beta_2} (\vex{r}).
\end{equation}
The vertex $\mathfrak{V}_{4}$ [Fig.~\ref{FigRules}(b) and Eq.~(\ref{Vertex})]
pairwise permutes the lower indices of composite operator ``legs,'' as shown in Fig.~\ref{FigScalingDim}(b). 
The completely antisymmetrized
operator defined by Eq.~(\ref{LDOSEigenOps}) is clearly an eigenoperator 
at one loop, since the sum of all diagrams to this order represents 
a complete
symmetrization procedure.
This 
is expected to hold to all higher orders in $t$, because 
$\mathcal{O}^{\; \alpha_{1} \ldots \alpha_{p}}_{p \; [\beta_{1} \ldots \beta_{p}]}$
plays the role of
a ``highest weight state''
in an irreducible representation of the full NL$\sigma$M target manifold symmetry group
$\mathrm{U}(2 n)$.

At one loop, the matrix element defined by Eq.~(\ref{MatrixElemDef}) 
is equal to
\begin{multline}\label{MatrixElem1loop}
	\Gamma^{(0)\phantom{p} \lambda_{1}\ldots\lambda_{p}}_{p \phantom{(0)}\gamma_{1}\ldots\gamma_{p}}
	[\mathcal{O}^{\; \alpha_{1} \ldots \alpha_{p}}_{p \; [\beta_{1} \ldots \beta_{p}]}(\vex{r})]
	\\
	\sim
	\left[1 - \frac{p(p-1)}{2} I_{1} \right] 
	\mathfrak{A}
	^{\phantom{p} \; \lambda_{1}\ldots\lambda_{p};\; \alpha_{1} \ldots \alpha_{p}}
	_{p \; \gamma_{1}\ldots\gamma_{p};\; [\beta_{1} \ldots \beta_{p}]},
\end{multline}	
where $\mathfrak{A}
	^{\phantom{p} \; \lambda_{1}\ldots\lambda_{p};\; \alpha_{1} \ldots \alpha_{p}}
	_{p \; \gamma_{1}\ldots\gamma_{p};\; [\beta_{1} \ldots \beta_{p}]}$ 
is the zeroth order amplitude [equal to zero or the pure constant $(1/p!)^2$,
depending upon the matrix element].\cite{footnote-TreeAmplitudeExplained}
In order to save writing wherever possible, from this place forward we will adopt the 
following shorthand notation: underlined vertices ($\underline{\Gamma}$), operators ($\underline{\mathcal{O}}$), 
and tree level matrix elements ($\underline{\mathfrak{A}}$) should be understood as possessing the appropriate 
set of indices, and all indices in a given equation are matched (in the appropriate order). With these 
conventions established, Eq.~(\ref{MatrixElem1loop}) may be rewritten compactly as
\begin{equation}\label{MatrixElem1loopCompact}
	\underline{\Gamma}^{(0)}_{p}
	[\underline{\mathcal{O}}_{p}(\vex{r})]
	\sim
	\left[1 - \frac{p(p-1)}{2} I_{1} \right] 
	\underline{\mathfrak{A}}_{p}.
\end{equation}	
In Eqs.~(\ref{MatrixElem1loop}) and (\ref{MatrixElem1loopCompact}),
\begin{align}\label{I1}
	I_{1} &= \frac{-t_{0}}{2} \int \frac{\dvex{k}}{(2\pi)^{d}} \, 
	\frac{1}{|\vex{k}|^{2} + h_{0} t_{0}}
	\nonumber\\
	&\sim 
	\frac{t}{4\pi}\left[\frac{1}{\epsilon} + \frac{1}{2}\ln\left(\frac{h t e^{\gamma}}{4 \pi \mu^{2}} \right)\right].
\end{align}
Here, we have used Eq.~(\ref{DimReg}) and standard dimensional regularization technology;
$\gamma$ denotes the Euler–Mascheroni constant.

To renormalize Eq.~(\ref{MatrixElem1loopCompact}), we insist that\cite{FieldTheory} 
\begin{equation}\label{CompOpRCD}
	Z_{p}^{-1} Z_{W}^{p/2} \underline{\Gamma}^{(0)}_{p}
	[\underline{\mathcal{O}}_{p}(\vex{r})] 
	= \textit{finite},
\end{equation}
where $Z_{p}$ is the composite operator renormalization, and the factor of 
$Z_{W}^{p/2}$ compensates for the $p$ (amputated) external fields. 
One then obtains the scaling dimension
\begin{align}\label{CompOpScalingDim}
	x_{p} &= \epsilon \frac{d\ln Z_{p}}{d\ln t}  
	\nonumber\\
	&= \frac{t}{4\pi}\left[n p - \frac{p(p-1)}{2}\right] + \ord{t^{2}}.
\end{align}

We make two observations.
First, by evaluating Eq.~(\ref{CompOpScalingDim}) at the Anderson transition critical point 
located by $t^{*} = 4\pi\sqrt{2\epsilon}$, with $n \rightarrow 0$, we recover the known 
results\cite{Pruisken85,Wegner87} for the multifractal spectrum $\tilde{\tau}(p)$ associated 
to the \emph{averaged} IPR, as provided above in Eqs.~(\ref{AvgMFCSpecI})--(\ref{XiDef}).
Second, we have only considered the renormalization of the fully antisymmetrized operator defined 
by Eq.~(\ref{LDOSEigenOps}), because this is the most relevant in the $n \rightarrow 0$ limit.
The fully \emph{symmetrized} operator 
$\mathcal{O}^{\; \alpha_{1} \alpha_{2} \ldots \alpha_{p}}_{p \; (\beta_{1} \beta_{2} \ldots \beta_{p})}$,
which is defined as in Eq.~(\ref{LDOSEigenOps}) without the $\sgn(\bm{\mathrm{P}})$ factor in the summand, 
also constitutes an eigenoperator with scaling dimension
\begin{align}\label{CompOpScalingDimSym}
	x_{p \, (\mathsf{sym})}
	= \frac{t}{4\pi}\left[n p + \frac{p(p-1)}{2}\right] + \ord{t^{2}}.
\end{align}
Consider the case of $n = 1$. The fully antisymmetrized operator defined by 
Eq.~(\ref{LDOSEigenOps}) does not exist for $p > 1$, since all $\Wh \Rightarrow W$ fields are 
scalars in this case.
At the fixed point located by $t^{*} = 4\pi\epsilon/n$ with $n = 1$, the symmetrized operator 
scaling dimension $x^{*}_{p \, (\mathsf{sym})} = \epsilon p(p+1)/2 + \ord{\epsilon^{2}}$, which 
is the expected result for spherical harmonic composite operators in the 
U(2)/U(1)$\times$U(1) $\sim$ O(3)/O(2) NL$\sigma$M.\cite{BrezinZinnJustinLeGuillou76}

\subsection{Two-point
function normalization at the MIT}\label{TECH3--Norm}

The set of coefficients $\{C_{q, q'}^{q+q'}\}$ [Eq.~(\ref{OPEDef})] defining the operator
product expansion (OPE) for \emph{properly normalized} composite eigenoperators 
constitute universal numbers characterizing the MIT. 
The proper (RG scheme-dependent) normalization of each eigenoperator is such that its two-point 
autocorrelation function is scheme-independent at the critical point in $d = 2 + \epsilon$.\cite{Cardy96} 
In this subsection, we derive the normalization of the operators defined by Eq.~(\ref{LDOSEigenOps}) 
with respect to their two-point functions (at large spatial separation), while the OPE is considered 
in the sequel. 
Since we are interested in critical properties, we assume $h = 0$ in Eq.~(\ref{SDef}) 
throughout the following discussion.

The technical tool for computing operator correlation functions at any perturbatively accessible
fixed point is RG-improved perturbation theory (PT). 
In the case of the non-trivial NL$\sigma$M critical point in $d = 2+\epsilon$, however, some technical 
difficulties arise:
For a NL$\sigma$M with a compact, non-Abelian symmetry, the trivial fixed point 
located at $t = 0$ is invariably infrared (IR)  unstable in 2D.\cite{FieldTheory} 
For any nonzero $t$, such a 2D model always flows under the RG toward a symmetry-restored, 
thermally-disordered ``paramagnetic'' state. 
Renormalized perturbation theory at $t \ll 1$, for composite operator correlators 
that are \emph{not} invariant under the full symmetry group of the 
sigma model target manifold, is typically plagued by IR divergences,
and hence affected by the specific way one regularizes these IR divergences.

The solution\cite{Elitzur,McKaneStone80,AmitKotliar80} that we employ in this section is to consider only invariant correlation 
functions. Invariant correlators are free of IR divergences, and a sensible renormalized PT for these objects can be 
constructed.\cite{footnote-IRinNLsM}
For the OPE in the next section, we will see that this is restriction is unnecessary.

We stress that, 
by the same token, 
\emph{all} eigenoperators at the non-trivial fixed point in $d = 2+\epsilon$ 
possess well-defined critical correlations.
Thus the above-described calculational impasse, as well as its solution, in fact reflect 
peculiarities of the $\epsilon$-expansion, rather than the NL$\sigma$M itself (at least for $d > 2$).

Within the symmetry-broken phase, `$\pi$' ($\Wh$,$\Wh^{\dagger}$) and `$\sigma$' 
($\sqrt{\hat{\mathbb{I}}_{n} - \Wh \Wh^{\dagger}}$,$\sqrt{\hat{\mathbb{I}}_{n} - \Wh^{\dagger} \Wh}$) 
fields possess very different correlation functions: the former constitute Goldstone modes with massless correlations, 
while the latter are gapped longitudinal modes, 
with massive correlation functions for all $t < t^{*}$.
At the critical point $t = t^{*}$ for $n > 0$, symmetry is restored; here, all operators belonging 
to a given irreducible representation of the target manifold symmetry group will possess identical 
correlations, provided a group-invariant normalization is chosen for these operators. [For the 
$\mathrm{O}(3)/\mathrm{O}(2)$ model, an invariant normalization is that conventionally assigned to 
spherical harmonics, written in terms of $\pi$ and $\sigma$ coordinates.] We will use this fact to 
determine the two-point function normalization of 
$\mathcal{O}^{\; \alpha_{1} \alpha_{2} \ldots \alpha_{p}}_{p \; [\beta_{1} \beta_{2} \ldots \beta_{p}]}$
[Eq.~(\ref{LDOSEigenOps})] for 
$n = \{1,2,\ldots\}$, and then continue the result to $n \rightarrow 0$.

Consider the following invariant,
``non-local'' operator,
\begin{equation}\label{InvariantOp}
	\Omega_{p}(\vex{r},\vex{r'}) \equiv 
	\sum_{\{m\}} \Theta_{p\, \{m\}}(\vex{r}) \Theta^{*}_{p\, \{m\}}(\vex{r'}),
\end{equation}
where $\Theta_{p\, \{m\}}$ is a composite operator that is a component of an irreducible representation
of the sigma model symmetry group. The representation is distinguished by the Casimir parameter $p$, while
the component operators are labeled 
by a set of ``magnetic'' quantum numbers $\{m\}$, 
e.g.\  
$\{\alpha_{1},\ldots,\alpha_{p},[\beta_{1},\ldots,\beta_{p}]\} \in \{m\}$ 
for the antisymmetrized LDOS 
moment operators defined by Eq.~(\ref{LDOSEigenOps}).
One may use  expressions
for the $\Theta_{p\, \{m\}}$ 
in terms of the target manifold coordinates 
$(\Wh,\Wh^\dagger,\sqrt{\hat{\mathbb{I}}_{n} - \Wh \Wh^{\dagger}},\sqrt{\hat{\mathbb{I}}_{n} - \Wh^{\dagger} \Wh})$ 
in order to construct an explicit expression for the RHS of
Eq.~(\ref{InvariantOp});
at the non-trivial critical point in $d = 2+\epsilon$, however, we require only
the lowest order expansion for the \emph{expectation} of Eq.~(\ref{InvariantOp}) in powers of 
$t^{*} \propto \epsilon^\sigma$ ($\sigma  = 1$ or $1/2$ for 
$n \in \mathbb{N}$ 
or $n\rightarrow0$, respectively). 
Since an expansion in powers of $t$ is equivalent to 
an expansion in powers of $\Wh$ and $\Wh^{\dagger}$, 
we make the following ansatz:
\begin{widetext}
\begin{equation}\label{InvariantOpExp}
	\Omega_{p}(\vex{r},\vex{r'})
	\sim
	\tilde{\alpha}_{p}
	\left\lgroup
	1 + \frac{\alpha_{p,1}}{2 n^{2}} 
	\Tr\left[
	\Wh^{\dagger}(\vex{r})\Wh(\vex{r'}) -
	\Wh^{\dagger}(\vex{r})\Wh(\vex{r}) 
	+
	\Wh^{\dagger}(\vex{r'})\Wh(\vex{r}) -
	\Wh^{\dagger}(\vex{r'})\Wh(\vex{r'})
	\right]
	+ \ord{\Wh^{4}}
	\right\rgroup.
\end{equation}
\end{widetext}
The expectation of the assumed form of Eq.~(\ref{InvariantOpExp}) is free from 
IR divergences.
While the numerical coefficient $\tilde{\alpha}_{p}$ in this equation is arbitrary, $\alpha_{p,1}$ is not, and can in principle
be computed from knowledge of the representation theory of the NL$\sigma$M symmetry group; instead,
we will determine its value empirically, below. Note that Eq.~(\ref{InvariantOpExp}) is
manifestly invariant under U($n$)$\times$U($n$) subgroup transformations, $\Wh \rightarrow \hat{U}_L \Wh \hat{U}_R$,
with 
$\hat{U}^{\dagger}_{L/R} \hat{U}^{\phantom{\dagger}}_{L/R} = \hat{\mathbb{I}}_{n}$.

We compute the expectation of Eq.~(\ref{InvariantOpExp})
in position space,\cite{McKaneStone80} using the IR-convergent Green's function at zero coupling 
\begin{multline}\label{IRGF}
	\langle 
	W^{\dagger\,\alpha_1}{}_{\beta_1}(\vex{r}) W^{\alpha_2}{}_{\beta_2}(0)
	- W^{\dagger\,\alpha_1}{}_{\beta_1}(0) W^{\alpha_2}{}_{\beta_2}(0)
	\rangle_{0}
	\\
	\begin{aligned}[b]
	&=
	\delta^{\alpha_1}_{\beta_2} \delta^{\alpha_2}_{\beta_1} \,
	t_{0} \int \frac{\dvex{k}}{(2\pi)^{d}} \,
	\frac{
	e^{i \vex{k}\cdot\vex{r}} - 1
	}{
	|\vex{k}|^2
	}
	\\
	&=
	\delta^{\alpha_1}_{\beta_2} \delta^{\alpha_2}_{\beta_1}
	\frac{t_{0}}{(d-2) S_{d}} 
        \frac{1}{|\vex{r}|^{d-2}}
	\\
	&\sim
	\delta^{\alpha_1}_{\beta_2} \delta^{\alpha_2}_{\beta_1}
	\frac{t}{2 \pi}
	\left[
	\frac{1}{\epsilon} - \frac{1}{2}
        \ln\left(\pi \mu^{2} |\vex{r}|^{2} e^{\gamma} \right)
	\right],
	\end{aligned}
\end{multline}
where $S_{d}$ is the surface area of the sphere in $d$ dimensions, 
and we have used Eqs.~(\ref{Prop}) and (\ref{DimReg}). 

Let us define the renormalized operator
\begin{equation}\label{InvariantOpRenorm}
	\left[\Omega_{p}\right](\vex{r},\vex{r'})
	\equiv 
	Z_{p}^{-2}
	\Omega_{p}(\vex{r},\vex{r'}),
\end{equation}
where $Z_{p}$ is the renormalization factor obtained via 
Eqs.~(\ref{CompOpRCD}) and (\ref{CompOpScalingDim}) for
the composite operators defined by Eq.~(\ref{LDOSEigenOps});
all operators belonging to a particular irreducible representation receive
the same renormalization in a NL$\sigma$M.\cite{BrezinZinnJustinLeGuillou76}

Insisting that $\langle \left[\Omega_{p}\right](\vex{r},\vex{r'}) \rangle$ is finite
(for $\vex{r} \neq \vex{r'}$), we see that we must take 
\begin{equation}
	\alpha_{p,1} = n p - \frac{p(p-1)}{2}
\end{equation}
in Eq.~(\ref{InvariantOpExp}). 
We have obtained the lowest order expansion coefficient for the group-invariant structure
defined by Eqs.~(\ref{InvariantOp}) and (\ref{InvariantOpExp}) without explicitly employing group theory, but using
only the renormalizability of the NL$\sigma$M! [Basic group theoretic knowledge was necessary to 
identify the invariant scaling operators defined by Eq.~(\ref{LDOSEigenOps}), however.]
\cite{footnote-Gegenbauer}

Finally, we set $t = t^{*}(\epsilon)$, and then 
we re-exponentiate the expectation of Eq.~(\ref{InvariantOpRenorm}) to obtain,
at the non-trivial critical point,
\begin{align}\label{InvariantOpCorr}
	\langle \left[\Omega_{p}\right](\vex{r},\vex{r'}) \rangle
	\sim 
	\frac
	{\tilde{\alpha}_{p} \left( \pi \mu^{2} e^{\gamma} \right)^{-x^{*}_{p}}}
	{|\vex{r}-\vex{r'}|^{2 x^{*}_{p}}},
\end{align}
where $x_{p}^{*}$ is the scaling dimension in Eq.~(\ref{CompOpScalingDim}),
evaluated at $t = t^{*}$. [$x_{p}^{*}$ is given explicitly 
by
Eqs.~(\ref{AvgMFCSpecxq})--(\ref{XiDef})
for the 
limit $n\rightarrow0$, appropriate to the MIT.]
 
Eq.~(\ref{InvariantOpCorr}) allows us to define the following ``renormalized and normalized'' 
composite operators, which we will enclose with the double square brackets $\NoOp{\cdots}$.
Referring to Eq.~(\ref{LDOSEigenOps}), we designate
\begin{equation}\label{LDOSEigenOpsNorm}
	\NoOp{
	\mathcal{O}^{\; \alpha_{1} \ldots \alpha_{p}}_{p \; [\beta_{1} \ldots \beta_{p}]}
	}
	(\vex{r})
	\equiv
	Z_{p}^{-1}
	\left( \pi \mu^{2} e^{\gamma} \right)^{\frac{x^{*}_{p}}{2}}
	\mathcal{O}^{\; \alpha_{1} \ldots \alpha_{p}}_{p \; [\beta_{1} \ldots \beta_{p}]}(\vex{r}).
\end{equation}
Eqs.~(\ref{InvariantOpRenorm}) and (\ref{InvariantOpCorr}) guarantee that 
the
two-point correlation function 
between distant operators defined by Eq.~(\ref{LDOSEigenOpsNorm}) is both ultraviolet finite, and 
independent of the renormalization scheme.

\subsection{Operator product expansion at the MIT}\label{TECH4--OPE}

We conclude this section with the construction of the operator product expansion (OPE) for the 
operators in Eq.~(\ref{LDOSEigenOpsNorm}). At the critical point in $d = 2 +\epsilon$, 
the OPE is expected to take the form
\begin{multline}\label{OPEDef2}
	\NoOp{
	\mathcal{O}^{\; \alpha_{1} \ldots \alpha_{p}}_{p \; [\beta_{1} \ldots \beta_{p}]}
	}(\vex{r})
	\;
	\NoOp{
	\mathcal{O}^{\; \alpha'_{1} \ldots \alpha'_{p'}}_{p' \; [\beta'_{1} \ldots \beta'_{p'}]}
	}(\vex{r'})
	\\
	\sim
	\frac{
	C_{p,p'}^{p+p'}
	}
	{
	|\vex{y}|^{x_{p}^{*}+x_{p'}^{*}-x_{p+p'}^{*}}
	}
	\NoOp{
	\mathcal{O}^{\; \alpha_{1} \ldots \alpha_{p} \alpha'_{1} \ldots \alpha'_{p'}}_{p+p' \; [\beta_{1} \ldots \beta_{p} \beta'_{1} \ldots \beta'_{p'}]}
	}(\vex{R})
	+ \ldots,
\end{multline}
where $\vex{y} \equiv \vex{r}-\vex{r'}$, $\vex{R} \equiv (\vex{r}+\vex{r'})/2$, $x_{p}^{*}$ is the scaling
dimension defined by Eqs.~(\ref{CompOpScalingDim}) and (\ref{AvgMFCSpecxq})--(\ref{XiDef}),
and $C_{p,p'}^{p+p'}$ is the (universal) OPE coefficient that we seek, expected to possess an 
expansion in powers of $t^{*}(\epsilon)$. Eq.~(\ref{OPEDef2}) will hold as a replacement rule 
in the limit $|\vex{y}| \rightarrow 0$, valid inside correlation functions involving arbitrary 
configurations of other, spatially remote operators. Note that Eq.~(\ref{OPEDef2}) relates a product 
of maximally antisymmetric operators to a single, maximally antisymmetric operator; the ellipsis ``$\ldots$'' 
on the right-hand side (RHS) of this equation represents other, less relevant operators that are 
produced in the `fusion' process; we will ignore the contribution of the latter to functional
RG.\cite{footnote-OPEIndexMatching}

In order to determine $C_{p,p'}^{p+p'}$, we compute an arbitrary matrix element 
$\Gamma^{\phantom{p+p'}\,\lambda_{1}\ldots\lambda_{p+p'}}_{p+p'\,\gamma_{1}\ldots\gamma_{p+p'}}[\cdots]$
of both sides of Eq.~(\ref{OPEDef2}), 
as defined by Eq.~(\ref{MatrixElemDef}). Using Eqs.~(\ref{MatrixElem1loop}) and (\ref{LDOSEigenOpsNorm}), 
and employing the compact notation introduced above and implemented in Eq.~(\ref{MatrixElem1loopCompact}),
the RHS of Eq.~(\ref{OPEDef2}) may be written as
\begin{align}\label{RHS1}
	\underline{\Gamma}_{p+p'}
	[\underline{\mathrm{RHS}}]
	=&
	Z_{W}^{\frac{p+p'}{2}} 
	Z_{p+p'}^{-1} 
	\left( \pi \mu^{2} e^{\gamma} \right)^{\frac{x^{*}_{p+p'}}{2}}
	\frac{
	C_{p,p'}^{p+p'}
	}
	{
	|\vex{y}|^{x_{p}^{*}+x_{p'}^{*}-x_{p+p'}^{*}}
	}
	\nonumber\\
	&\times
	\underline{\Gamma}^{(0)}_{p+p'}[\underline{\mathcal{O}}_{p+p'}(\vex{R})],
\end{align} 
where $Z_{W}$ ($Z_{p+p'}$) is the field strength (composite operator) renormalization
factor, and the bare amplitude is given by
\begin{align}\label{RHS2}
	\underline{\Gamma}^{(0)}_{p+p'}[\underline{\mathcal{O}}_{p+p'}]
	= 
	\left[1 - \frac{1}{2}(p+p')(p+p'-1) I_{1}\right] \underline{\mathfrak{A}}_{p+p'}.
\end{align}
In Eq.~(\ref{RHS2}), $I_{1}$ is the loop integral defined by Eq.~(\ref{I1}), where we retain the 
infrared regularization for now, $h_{0} \neq 0$, while 
\[	
	\underline{\mathfrak{A}}_{p+p'}\rightarrow 
	\mathfrak{A}
	^{\phantom{p+p'}\,\lambda_{1}\ldots\lambda_{p+p'};\; \alpha_1 \ldots \alpha_{p} \alpha'_1 \ldots \alpha'_{p'}}
	_{p+p'\,\gamma_{1}\ldots\gamma_{p+p'};\; [\beta_1 \ldots \beta_{p} \beta'_1 \ldots \beta'_{p'}]}
\]	
denotes the zeroth order
amplitude for the matrix element (a pure number).
(See footnote~\onlinecite{footnote-TreeAmplitudeExplained} for details.)  

Similarly, the LHS matrix element of Eq.~(\ref{OPEDef2}) may be written
\begin{align}\label{LHS1}
	\underline{\Gamma}_{p+p'}[\underline{\mathrm{LHS}}]
	=&
	Z_{W}^{\frac{p+p'}{2}} 
	Z_{p}^{-1} 
	Z_{p'}^{-1} 
	\left( \pi \mu^{2} e^{\gamma} \right)^{\frac{x^{*}_{p}+x^{*}_{p'}}{2}}
	\nonumber\\
	&\times
	\underline{\Gamma}^{(0)}_{p+p'}
	[\underline{\mathcal{O}}_{p}(\vex{r})\underline{\mathcal{O}}_{p'}(\vex{r'})],
\end{align}
where the bare (unrenormalized) amplitude is
\begin{subequations}
\begin{align}
	\underline{\Gamma}^{(0)}_{p+p'}&
	[\underline{\mathcal{O}}_{p}(\vex{r})\underline{\mathcal{O}}_{p'}(\vex{r'})]
	\nonumber\\
	\rightarrow&
	\begin{aligned}[t]
	\langle\langle 
	\mathcal{O}^{\; \alpha_{1} \ldots \alpha_{p}}_{p \; [\beta_{1} \ldots \beta_{p}]}(\vex{r})\,&
	\mathcal{O}^{\; \alpha'_{1} \ldots \alpha'_{p'}}_{p' \; [\beta'_{1} \ldots \beta'_{p'}]}(\vex{r'})
	\\
	&
	\times W^{\dagger\, \lambda_{1}}{}_{\gamma_{1}} \cdots W^{\dagger\, \lambda_{p+p'}}{}_{\gamma_{p+p'}}
	\rangle\rangle
	\end{aligned}
	\label{LHSBareA}\\
	=&
	\begin{aligned}[t]
	\frac{1}{(p! p'!)^{2}}
	\langle\langle
	&
	W^{\alpha_{1}}{}_{\beta_{1}} \cdots W^{\alpha_{p}}{}_{\beta_{p}}(\vex{r})
	\\
	&\times
	W^{\alpha'_{1}}{}_{\beta'_{1}} \cdots W^{\alpha'_{p'}}{}_{\beta'_{p'}}(\vex{r'})
	\\
	&\times 
	W^{\dagger\, \lambda_{1}}{}_{\gamma_{1}} \cdots W^{\dagger\, \lambda_{p+p'}}{}_{\gamma_{p+p'}}
	\rangle\rangle
	\end{aligned}
	\nonumber\\
	&+  \;
	\left\{
	\begin{aligned}
	&(p! p'!-1) \; \textit{other terms}
	\\
	&\textit{obtained by permutations}
	\end{aligned}
	\right\}\label{LHSBareB}.
\end{align}
\end{subequations}
As in Eq.~(\ref{MatrixElemDef}), the external fields $\{\Wh^{\dagger}\}$ in these equations 
are assumed to be located far from the vicinity of the operator product ($\vex{r}$ or $\vex{r'}$) 
and from each other, while the double angle brackets instruct us to take the one-particle irreducible amplitude, 
with external legs amputated.\cite{FieldTheory}

\begin{figure}
\includegraphics[width=0.48\textwidth]{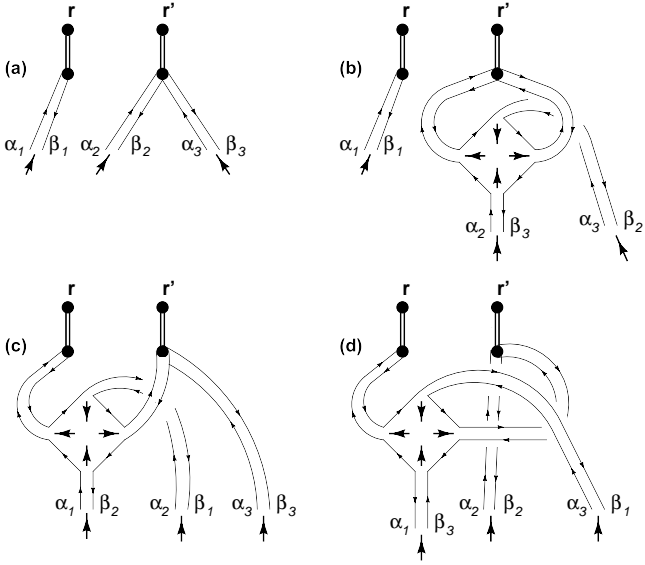}
\caption{Basic renormalization process in the OPE.
\label{FigOPE}}
\end{figure}

Consider the one-loop renormalization of the term written explicitly in Eq.~(\ref{LHSBareB}); the other $(p!p'! - 1)$
terms implied in this equation will give identical contributions. The basic renormalization process of an operator
product is illustrated in Fig.~\ref{FigOPE} specifically for the combination 
\begin{equation}
	W^{\alpha_{1}}{}_{\beta_{1}}(\vex{r}) \,\otimes\, 
	W^{\alpha_{2}}{}_{\beta_{2}}(\vex{r'})
	W^{\alpha_{3}}{}_{\beta_{3}}(\vex{r'}).
	\nonumber
\end{equation}
The vertex $\mathfrak{V}_{4}$ [Fig.~\ref{FigRules} and Eq.~(\ref{Vertex})] modifies the operator product in two ways.
First, it renormalizes the constituent operators, pairwise permuting indices of legs both tied to either $\vex{r}$ or $\vex{r'}$, 
as shown in Fig.~\ref{FigOPE}(b) [c.f.\ Fig.~\ref{FigScalingDim}]. Second, $\mathfrak{V}_{4}$ ties the two operators 
together by pairwise crosspermuting their indices, as depicted in Figs.~\ref{FigOPE}(c) and (d). 

Now, algebraically we may express an unsymmetrized product of $q$ 
$\hat{W}$-field matrix elements 
$(W^{\alpha}{}_{\beta})$ in terms of the completely 
antisymmetrized product, native to the most relevant
irreducible representation with Casimir parameter $q$ of the NL$\sigma$M symmetry group, 
plus other terms which belong to other 
(completely symmetric or mixed symmetry) representations. In particular,
\begin{equation}\label{qSym}
	W^{\alpha_{1}}{}_{\beta_{1}} \cdots W^{\alpha_{q}}{}_{\beta_{q}}
	= W^{\alpha_{1}}{}_{[\beta_{1}} \cdots W^{\alpha_{q}}{}_{\beta_{q}]} + \ldots,
\end{equation}
where again the square brackets $[\cdots]$ denote complete antisymmetrization. 
The unity coefficient in front of the completely 
antisymmetrized tensor on the RHS of Eq.~(\ref{qSym}) follows from the fact that the 
antisymmetrization procedure is a projective operation that kills 
symmetric or mixed symmetry terms, but leaves the 
pre-antisymmetrized component invariant. 

After the vertex acts upon the unsymmetrized operator product displayed
explicitly in Eq.~(\ref{LHSBareB}), giving the appropriate factors for
the two types of renormalization depicted in
Fig.~\ref{FigOPE}(b) and 
Figs.~\ref{FigOPE}(c),(d), 
respectively, we are free to use Eq.~(\ref{qSym})
to replace each resulting unsymmetrized, permuted product with the corresponding completely 
antisymmetrized version, up to less relevant mixed symmetry or higher gradient terms.
\cite{footnote-highergrad}
The completely 
antisymmetrized
product of $p + p'$ factors is just the composite operator
on the RHS of the OPE, as defined by Eq.~(\ref{OPEDef2}). Therefore, using the Feynman rules
in Eqs.~(\ref{Prop}) and (\ref{Vertex}), and summing all diagram topologies to one loop,
Eq.~(\ref{LHSBareA}) may be written as
\begin{align}\label{LHSBareC}
	\underline{\Gamma}^{(0)}_{p+p'}&
	[\underline{\mathcal{O}}_{p}(\vex{r})\underline{\mathcal{O}}_{p'}(\vex{r'})]
	\nonumber\\
	\sim& 
	\binom{p+p'}{p}\!\!
	\left\lgroup
	\begin{aligned}
	1 
	&- I_{1}
	\left[
	\frac{p(p-1)}{2} + \frac{p'(p'-1)}{2}
	\right]
	\\
	&- I_{2}(\vex{y})
	\left[
	p p'
	\right]
	\end{aligned}
	\right\rgroup
	\underline{\mathfrak{A}}_{p+p'},
\end{align}
where $\underline{\mathfrak{A}}_{p+p'}$ denotes the zeroth order matrix element of 
	$
	\mathcal{O}
	^{\; \alpha_{1} \ldots \alpha_{p} \alpha'_{1} \ldots \alpha'_{p'}}
	_{p+p' \; [\beta_{1} \ldots \beta_{p} \beta'_{1} \ldots \beta'_{p'}]}
	$,
as in Eq.~(\ref{RHS2}), 
$I_{1}$ is the integral defined by Eq.~(\ref{I1}), and
\begin{align}\label{I2}
	I_{2}(\vex{y}) &= \frac{-t_{0}}{2} \int \frac{\dvex{k}}{(2\pi)^{d}} \, 
	\frac{e^{i \vex{k}\cdot\vex{y}}}{|\vex{k}|^{2} + h_{0} t_{0}}.
\end{align}
Let us briefly comment upon the origin of the various combinatoric factors in Eq.~(\ref{LHSBareC}):
The prefactor $\binom{p+p'}{p}$ originates 
from the
normalization convention used in Eq.~(\ref{LDOSEigenOps}). 
The factors of $p(p-1)/2$ and $p'(p'-1)/2$ count the number 
of inequivalent ways leg indices associated with \emph{either} operator 
$\underline{\mathcal{O}}_{p}$ or $\underline{\mathcal{O}}_{p'}$ 
(but not both) may be permuted, as occured previously in the scaling
dimension calculation [Eq.~(\ref{MatrixElem1loop})]. 
The factor of $p p'$ counts the number of 
inequivalent ways one leg index from each operator may be interpermuted, as in Figs.~\ref{FigOPE}(c),(d).

Equating the left-hand and right-hand sides of the OPE, Eqs.~(\ref{RHS1})--(\ref{LHS1}) 
and (\ref{LHSBareC}), and expanding everything to the lowest non-trivial order in $t$, we obtain
\begin{equation}\label{OPECoeff1}
	C_{p,p'}^{p+p'}
	\sim
	\binom{p+p'}{p}\!\!
	\left\lgroup
	\begin{aligned}
	&1 
	+ p p' \left[I_{1}-I_{2}(\vex{y})\right]
	+ \ln \left[\frac{Z_{p+p'}}{Z_{p} Z_{p'}}\right]	
	\\
	&- \frac{1}{2}(x_{p+p'}^{*}-x_{p}^{*}-x_{p'}^{*}) 
	\\
	&\phantom{-}
	\times\ln\left(\pi \mu^{2} |\vex{y}|^{2} e^{\gamma} \right)
	\end{aligned}
	\right\rgroup.
\end{equation}
We may now take the limit $h_{0} \rightarrow 0$, because the combination 
$2[I_{1}-I_{2}(\vex{y})]$ [Eqs.~(\ref{I1}) and (\ref{I2})] gives the IR-finite integral evaluated 
previously in Eq.~(\ref{IRGF}), above. Using Eq.~(\ref{CompOpScalingDim}), Eq.~(\ref{OPECoeff1})
simplifies to the expression
\begin{equation}\label{OPECoeff2}
	C_{p,p'}^{p+p'}
	\sim
	\binom{p+p'}{p}\!\!
	\left\lgroup
	\begin{aligned}
	1 
	- \frac{1}{2}&\left(x_{p+p'}^{*}-x_{p}^{*}-x_{p'}^{*} + t\frac{p p'}{4\pi}\right)
	\\
	&\times\ln\left(\pi \mu^{2} 
        |\vex{y}|^{2} e^{\gamma} \right)
	\end{aligned}
	\right\rgroup.
\end{equation}
In general, the RHS of this equation is a UV-finite, non-zero function of the operator
separation $\vex{y}$. At the non-trivial critical point $t = t^{*}$ in $d = 2+ \epsilon$,
however, we have [from Eq.~(\ref{CompOpScalingDim})]
\begin{equation}\label{OPECoeff3}
	C_{p,p'}^{p+p'}
	\sim
	\binom{p+p'}{p}
	+ \ord{t^{* \, 2}}.
\end{equation}
At the Anderson metal-insulator transition ($n \rightarrow 0$), $t^{*} = 4\pi\sqrt{2\epsilon} + \ord{\epsilon}$.
Thus the OPE coefficient for the operators with normalization determined by Eq.~(\ref{LDOSEigenOpsNorm})
is \emph{independent} of $\sqrt{\epsilon}$ to order $\epsilon$.\cite{comparison with BDI}

The result in Eq.~(\ref{OPECoeff3}) should be contrasted with a similar computation 
in $\phi^{4}$ theory in $d = 4 - \epsilon$: at the Wilson-Fisher fixed point, the fusion 
of two elementary
renormalized and normalized $\NoOp{\phi}$ fields into the mass operator 
$\NoOp{\phi^{2}}$ yields an OPE coefficient 
that acquires a correction at the lowest non-trivial order in the quartic coupling 
strength $\lambda^{*}\propto\epsilon + \ord{\epsilon^{2}}$. (See, e.g., Ref.~\onlinecite{Brown}.) 
As in the above NL$\sigma$M calculation, the normalization of the operators $\NoOp{\phi}$ and
$\NoOp{\phi^2}$ is chosen so as to give two-point autocorrelation functions independent of
the renormalization scheme.

Finally, 
tracing over pairs of indices in Eq.~(\ref{OPEDef2}) allows the OPE to be written as
\begin{equation}\label{OPEFinal}
	\NoOp{
	\mathcal{O}_{p}
	}(\vex{r})
	\;
	\NoOp{
	\mathcal{O}_{p'}
	}(\vex{r'})
	\sim
	\frac{
	C_{p,p'}^{p+p'}
	}
	{
	|\vex{y}|^{x_{p}^{*}+x_{p'}^{*}-x_{p+p'}^{*}}
	}
	\NoOp{
	\mathcal{O}_{p+p'}
	}(\vex{R})
	+ \ldots,
\end{equation}
where
\begin{align}\label{LDOSEigenOpsSummedNorm}
	\NoOp{\mathcal{O}_{p}}(\vex{r})
	& \equiv
	\sum_{\alpha_{1} = 1}^{n}\ldots\sum_{\alpha_{p} = 1}^{n}
	\NoOp{
	\mathcal{O}^{\; \alpha_{1} \ldots \alpha_{p}}_{p \; [\alpha_{1} \ldots \alpha_{p}]}
	}(\vex{r})
	\nonumber\\
	&=
	Z_{p}^{-1}
	\left( \pi \mu^{2} e^{\gamma} \right)^{\frac{x^{*}_{p}}{2}}
	\mathcal{O}_{p}(\vex{r})
\end{align}
is the renormalized and normalized version of the bare operator defined by 
Eq.~(\ref{LDOSEigenOpsSummed}).

The OPE in Eq.~(\ref{OPEFinal}), together with the coefficient $C_{p,p'}^{p+p'}$ 
given by Eq.~(\ref{OPECoeff3}), constitutes the primary technical result of this paper.
We have succeeded in associating a unique, properly normalized operator $\NoOp{\mathcal{O}_{p}}(\vex{r})$
to the $p^{\mathrm{th}}$ moment of the LDOS, and demonstrated that the family of such
operators obeys the OPE set forth in Eq.~(\ref{OPEDef}) in the Introduction.
This is the necessary input to the functional RG scheme used in Sec.~\ref{sec: FRG} to
extract the typical multifractal spectrum $\tau(q)$ in Eq.~(\ref{tau(q)typFinalC}).

\begin{acknowledgments}

We would like to thank Igor Aleiner, Boris Altshuler, and Gabriel Kotliar
for helpful discussions and probing questions.
This work was supported in part by the Nanoscale Science and Engineering Initiative of the
National Science Foundation under NSF Award Number CHE-06-41523, and by the New York State Office 
of Science, Technology, and Academic Research (NYSTAR) (M.S.F.).
This work was also supported in part by the NSF under Grants No.\ 
DMR-0547769 (M.S.F.),
PHY05-51164 (S.R.), and DMR-0706140 (A.W.W.L.).
S.R.\ thanks the Center for Condensed Matter Theory at University of 
California, Berkeley for its support.

\end{acknowledgments}



\begin{thebibliography}{99}
\bibitem{x1For2DchiralDirac}
	In contrast,
	among the non-standard classes\cite{DisClasses} extensively studied 
	in the recent past, including the 2D Dirac models
	\cite{Ludwig94,Mudry96,Kogan96,Chamon96,Castillo97,Carpentier00,Carpentier01,Motrunich02,Mudry03,Fukui02,Yamada03}
	that inspired the present work, the average density of states may
	alternatively vanish or diverge at the delocalization transition,
	implying that $x^*_{1} \neq 0$.  Eq.~(\ref{AvgMFCSpecI})
	applies equally to this more general case.
\bibitem{symplecticClass}
	Although the results for 
	the symplectic symmetric class are also known,
	they have no direct physical application in Anderson localization.
	There is no perturbatively accessible, infra-red non-trivial fixed point
	in the symplectic symmetry class once the replica limit is
	taken.
\bibitem{footnote-Clebsch}
	The key to Eq.~(\ref{O(3)EigenOpA}) is the standard angular momentum addition formula
	\begin{multline}
		Y_{l_1,m_1}(\pi_{+},\pi_{-},\sigma) \, Y_{l_2,m_2}(\pi_{+},\pi_{-},\sigma) 
		\\= 
		\sum_{l,m} 
		\, \langle l_1 l_2; l m| l_1 l_2 ; m_1 m_2\rangle
		\, c^{l}_{l_1,l_2} 
		\, Y_{l,m}(\pi_{+},\pi_{-},\sigma),
		\nonumber
	\end{multline}
	where $\langle l_1 l_2; l m| l_1 l_2 ; m_1 m_2\rangle$ is a Clebsch-Gordan coefficient,
	and $c^{l}_{l_1,l_2}$ is a normalization constant that is independent of $\{m,m_1,m_2\}$.
	As an example, one has
	\[
		(\pi_{+} + \pi_{-})^2 = 
		\frac{1}{\lambda_{2}}\left[Y_{2,2} - \sqrt{\frac{2}{3}}Y_{2,0} + Y_{2, -2} \right]
		+\frac{1}{\lambda_{0}} \frac{4}{3} Y_{0,0},
	\]
	where the coefficients $\{\lambda_{2},\lambda_{0}\}$ are determined by Eq.~(\ref{lambdaNormDef}).
\bibitem{footnote-OtherEigenOps}
	Because of the symmetry restoration linking the `$\pi$' ($\hat{W},\hat{W}^{\dagger}$) 
	and `$\sigma$' ($\sqrt{\hat{\mathbb{I}}_{n} - \hat{W}\hat{W}^{\dagger}},\sqrt{\hat{\mathbb{I}}_{n} - \hat{W}^{\dagger}\hat{W}}$)
	NL$\sigma$M fields at the non-trivial critical point in $d = 2 + \epsilon$
	(for fixed $n \in \{1,2,\ldots\}$),
	many other eigenoperators exist that share the same ``maximally'' relevant eigenvalues as those 
	defined by Eqs.~(\ref{LDOSEigenOps}) and (\ref{LDOSEigenOpsSummed}). Our choice to focus upon the 
	latter operator family is one of convenience.
\bibitem{Yudson}
	A
	multi-local field theory,
	which can directly treat the probability distribution
	of the Green functions and the conductance, 
	was proposed by Yudson in Ref.~\onlinecite{Yudson05}.
\bibitem{footnote-VertexDef}
	Strictly speaking, the (bare, unrenormalized) matrix element in Eq.~(\ref{MatrixElemDef}) is a mixed 
	position and momentum space object, the precise definition of which is given by the correlation function
	\begin{multline}\label{MatrixElemDefPrecise}
	\Gamma^{(0)\phantom{p} \lambda_{1}\ldots\lambda_{p}}_{p \phantom{(0)}\gamma_{1}\ldots\gamma_{p}}
	[\mathcal{O}^{\; \alpha_{1} \ldots \alpha_{p}}_{p \; [\beta_{1} \ldots \beta_{p}]}(\vex{r})]
	\\
	= 
	\frac{
	\langle 
	\mathcal{O}^{\; \alpha_{1} \ldots \alpha_{p}}_{p \; [\beta_{1} \ldots \beta_{p}]}(\vex{r})\,\,
	W^{\dagger\, \lambda_{1}}{}_{\gamma_{1}}(\vex{k}_{1}) \cdots W^{\dagger\, \lambda_{p}}{}_{\gamma_{p}}(\vex{k}_{p})
	\rangle
	}
	{
	G(\vex{k}_{1}) G(\vex{k}_{2}) \cdots G(\vex{k}_{p})
	},
	\end{multline}
	where $G(\vex{k})$ is the ``one-particle'' propagator, which has the amplitude given
	by Eq.~(\ref{Prop}) at tree level. Only completely connected diagrams are counted in the 
	evaluation of the right-hand side of Eq.~(\ref{MatrixElemDefPrecise}). The 
	external momenta $\{\vex{k_{1}},\ldots,\vex{k}_{p}\}$ can be taken to vanish
	in this equation.
\bibitem{footnote-TreeAmplitudeExplained}
	The tree level amplitude 
	$\mathfrak{A}
	^{\phantom{p} \; \lambda_{1}\ldots\lambda_{p};\; \alpha_{1} \ldots \alpha_{p}}
	_{p \; \gamma_{1}\ldots\gamma_{p};\; [\beta_{1} \ldots \beta_{p}]}$ 
	can be written as a sum of $2p$-fold products of Kronecker delta functions.
	In the case of the operator
	$
	\mathcal{O}^{\; \alpha_{1} \alpha_{2}}_{2 \; [\beta_{1} \beta_{2}]}(\vex{r})
	=
	(1/2!)^2
	[
	W^{\alpha_{1}}{}_{\beta_{1}} W^{\alpha_{2}}{}_{\beta_{2}} (\vex{r})
	-
	W^{\alpha_{1}}{}_{\beta_{2}} W^{\alpha_{2}}{}_{\beta_{1}} (\vex{r})
	]
	$,
	one finds the amplitude
	$\mathfrak{A}
	^{\phantom{2} \; \lambda_{1} \lambda_{2};\; \alpha_{1} \alpha_{2}}
	_{2 \; \gamma_{1} \gamma_{2};\; [\beta_{1} \beta_{2}]}
	= 
	(1/2!)^2
	[
	\delta^{\lambda_{1}}_{\beta_{1}} \delta_{\gamma_{1}}^{\alpha_{1}}
	\delta^{\lambda_{2}}_{\beta_{2}} \delta_{\gamma_{2}}^{\alpha_{2}}
	+
	\delta^{\lambda_{1}}_{\beta_{2}} \delta_{\gamma_{1}}^{\alpha_{2}}
	\delta^{\lambda_{2}}_{\beta_{1}} \delta_{\gamma_{2}}^{\alpha_{1}}
	-
	\delta^{\lambda_{1}}_{\beta_{2}} \delta_{\gamma_{1}}^{\alpha_{1}}
	\delta^{\lambda_{2}}_{\beta_{1}} \delta_{\gamma_{2}}^{\alpha_{2}}
	-
	\delta^{\lambda_{1}}_{\beta_{1}} \delta_{\gamma_{1}}^{\alpha_{2}}
	\delta^{\lambda_{2}}_{\beta_{2}} \delta_{\gamma_{2}}^{\alpha_{1}}
	]
	$.
\bibitem{footnote-IRinNLsM}
	As discussed in Ref.~\onlinecite{AmitKotliar80}, the basic problem is that non-invariant correlation 
	functions suffer ultraviolet (UV) divergences for $d \geq 2$, and infrared (IR) divergences for 
	$d \leq 2$---there is no finite window of dimensionality free from both UV and IR problems. Within 
	dimensional regularization, non-covariant correlators cannot be regularized via analytical continuation 
	in $d$, so renormalized PT does not exist for these objects. By contrast, invariant correlators 
	can be shown to be IR and UV finite for $0 < d < 2$; in this case, dimensional regularization works
	in the usual way.
\bibitem{footnote-Gegenbauer}
	As an amusing aside, the same technique employed above can be used to derive the 
	Gegenbauer polynomials, expanded in terms of $\vec{\pi}$ and 
	$\sigma = 1 -\vec{\pi}\cdot\vec{\pi}/2 + \ldots$, in the O($n$)/O($n-1$) model, given a knowledge
	only of the one-loop scaling dimension for composite eigenoperators belonging to
	a particular irreducible representation.
\bibitem{footnote-OPEIndexMatching}
	Note that in order for the most relevant process in Eq.~(\ref{OPEDef2}) to occur, it is necessary
	that all of the lower indices and all of the upper indices appearing on the LHS of this 
	equation separately differ (i.e., no upper indices have the same value, no lower indices 
	have the same value, but upper indices may take the same values as lower ones). Otherwise, 
	the right-hand side of this equation vanishes, up to less relevant contributions.
\bibitem{footnote-highergrad}
	We neglect higher gradient operators throughout this paper, which are 
	important to the computation of conductance, rather than LDOS 
	fluctuations.\cite{AltshulerKratsovLerner86-89,AltshulerKratsovLerner91,Wegner91}
\end{thebibliography}
\end{document}